\documentclass{aa}

\usepackage{txfonts}
\usepackage[english]{babel}

\usepackage{graphicx}
\graphicspath{{Figures/}}

\usepackage{natbib}
\usepackage{xspace}
\usepackage{enumitem} 
\usepackage{color}
\usepackage{caption}
\usepackage{hyperref}
\usepackage{amsmath}
\usepackage{multirow}
\usepackage{amssymb, amsfonts}
\usepackage{soul}
\usepackage{comment}
\usepackage{graphicx}
\usepackage[caption=false]{subfig}

\usepackage{natbib}
\bibpunct{(}{)}{;}{a}{}{,}

\setlist[itemize]{noitemsep} 

\newcommand{\pcal}{\mathcal P}
\newcommand{\lcal}{\mathcal L}

\begin{document}
\title{\texttt{COVMOS}: a new Monte Carlo approach for galaxy clustering analysis}

\author{ Philippe Baratta\inst{1}
\and Julien Bel \inst{2}
\and Sylvain Gouyou Beauchamps \inst{1,3,4}
\and Carmelita Carbone \inst{5}
}

\institute{
Aix Marseille Universit\'e, CNRS/IN2P3, CPPM, IPhU, Marseille, France 
\and Aix Marseille Univ, Universit\'e de Toulon, CNRS, CPT, Marseille, France
\and Institute of Space Sciences (ICE, CSIC), Campus UAB, Carrer de Can Magrans, s/n, 08193 Barcelona, Spain
\and Institut d’Estudis Espacials de Catalunya (IEEC), Carrer Gran Capitá 2-4, 08034 Barcelona, Spain
\and Istituto di Astrofisica Spaziale e Fisica cosmica Milano, Via A. Corti 12, I-20133 Milano, Italy
}

\offprints{\mbox{P.~Baratta}, \email{baratta@cppm.in2p3.fr} }

\abstract{We validate the \texttt{COVMOS} method introduced in \cite{Baratta:2019bta} allowing for the fast simulation of catalogues of different cosmological field tracers (e.g. dark matter particles, halos, galaxies, etc.). The power spectrum and one-point probability distribution function of the underlying tracer density field are set as inputs of the method and are arbitrarily chosen by the user. In order to evaluate the validity domain of \texttt{COVMOS} at the level of the produced two-point statistics covariance matrix, we choose to target these two input statistical quantities from realistic $N$-body simulation outputs. In particular, we perform this cloning procedure in a $\Lambda$CDM and in a massive neutrino cosmologies, for five redshifts in the range $z\in[0,2]$. First, we validate the output real-space two-point statistics (both in configuration and Fourier space) estimated over $5,000$ \texttt{COVMOS} realisations per redshift and per cosmology, with a volume of $1\ [\mathrm{Gpc}/h]^3$ and $10^8$ particles each. Such a validation is performed against the corresponding $N$-body measurements, estimated from 50 simulations. We find the method to be valid up to $k\sim 0.2h/$Mpc for the power spectrum and down to $r~\sim 20$ Mpc$/h$ for the correlation function. Then, we extend the method by proposing a new modelling of the peculiar velocity distribution, aiming at reproducing the redshift-space distortions both in the linear and mildly non-linear regimes. After validating this prescription, we finally compare and validate the produced redshift-space two-point statistics covariance matrices in the same range of scales. We release on a public repository the Python code associated with this method, allowing the production of tens of thousands of realisations in record time. \texttt{COVMOS} is intended for any user involved in large galaxy-survey science requiring a large number of mock realisations.}


\maketitle

\section{Introduction}
Understanding the dark sector of the universe constitutes the main challenge of future large galaxy surveys like Euclid \citep{Laureijs:2011gra},  LSST \citep{Abell:2009aa} or DESI \citep{Aghamousa:2016zmz}. These have been designed to provide a large amount of data able to disentangle and constrain with high accuracy several cosmological parameters, as e.g. the dark energy equation of state, the neutrino mass, as well as more alternative cosmological scenarios, through galaxy clustering, weak lensing and their cross-correlation.

However, even with infinitely precise and voluminous data, discriminating among different cosmological models is not guaranteed as it also requires the development of specific analysis methods. The complexity of galaxy survey data analyses mainly lies on the fact that we should combine several types of correlated cosmological probes and identify many systematic errors and biases. Thus, such analyses must lay on robust statistical tools including a good estimation of correlations in the data and expected error bars.
In particular, the cosmological model extraction is based on the evaluation of a Likelihood, requiring an accurate covariance matrix for the considered observables \citep{Taylor:2012kz}.

Although predicting the galaxy clustering covariance in a fully analytical way \citep{wadekar_19, lacasa_17, krause_16} is possible, when one needs to cross-correlate various observables, alternatives to analytical covariances must be developed. As a matter of fact, most of the covariance matrices exploited in galaxy survey analysis today rely large samples of cosmological simulations \citep{Kitaura:2015uqa,zhao_21}, and even when an analytical covariance can be used, such as in the recent DES Y3 results \citep{friedrich_20}, this has to be validated against a covariance sampled from accurate numerical simulations. 

In addition to the fact that the construction of such large samples of mock catalogues is highly CPU and time consuming, the major problem is that the covariance is model-dependent: at least one covariance matrix per tested cosmological model is needed (when assuming in the analysis a covariance matrix constant with respect to the parameters to infer).

Different approaches have been investigated to bypass these issues, called approximate methods.
A category of them is designed to speed up the $N$-body evolution by either approximating the dynamics of the global dark matter field or by focusing on the evolution of early identified halos such as ICE-COLA \citep{Tassev_2013,Izard:2017kma}, PINOCCHIO \citep{Monaco:2001jg,10.1093/mnras/stt907} or PEAK PATCH \citep{1996ApJS..103....1B}). Although highly competitive with respect to $N$-Body simulations \citep[at least up to four-point statistics, see][]{Lippich:2018wrx, Blot:2018oxk, Colavincenzo:2018cgf}, they still remain CPU and time-demanding. 
Other types of methods, like PATCHY \citep{10.1093/mnrasl/slt172,10.1093/mnras/stv645} or HALOGEN \citep{Avila:2014nia} are faster than the above methods since they do not require any field evolution. They directly produce the dark matter field at the considered redshift, before applying a biasing scheme (using second order Lagrangian perturbation theory fitted to $N$-body products to obtain halo catalogues. However, these methods require a certain level of calibration and therefore remain indirectly expensive to adapt from one cosmological model to another.

Finally, a third category of approximate methods has already been studied in the specific case of Log-Normal density fields \citep[e.g.][]{Coles:1991if,Greiner:2013jea}. These methods consist in targeting, at a given redshift, both the clustering amplitude, usually through the power spectrum, and the probability distribution function (PDF) of the simulated density field. A discretisation of this field, based on the so-called Local Poisson process approximation, can eventually be applied to obtain a discrete set of objects. Because of the rapidity and the simplicity of implementation of such methods, a vast amount of literature has investigated the Log-Normal distributions \citep[e.g.][]{Xavier:2016elr,Alonso:2014sna, Agrawal:2017}. However it has been shown by \cite{Carron:2011et} that the Log-Normal statistics poorly reproduce the true data correlations when entering the non-linear regime, due to the limited range of validity of such approximated PDF forms. 

In this context we introduced in \cite{Baratta:2019bta} a new Monte-Carlo method fully suited for simulating catalogues of objects defined by an arbitrary power spectrum and a $1$-point PDF. 
In the present work, we aim at validating this method against the \texttt{DEMNUni} \citep[Dark Energy and Massive Neutrino Universe;][]{carbone_16, castorina_15, parimbelli_22} set of $N$-body simulations. 
We thus evaluate the validity domain of the $2$-point statistic covariance matrices produced with the \texttt{COVMOS}\footnote{Publicly available in  \href{https://github.com/PhilippeBaratta/COVMOS}{github.com/PhilippeBaratta/COVMOS}} public code. 

This paper is organised as follows. In Sec.~\ref{pipelinecomovingspace} we detail the core of the method and generalise it to the simulation of power spectra with non-linear prescriptions.
In Sec.~\ref{resultscomovingspace}, after introducing the \texttt{DEMNUni} $N$-Body simulations, we compare and discuss the results for the produced $2$-point statistics (the power spectrum and the $2$-point correlation function) in real space. We then extend the Monte Carlo method by introducing our peculiar velocity assignment procedure in Sec.~\ref{pipelineredshiftspace}, reproducing redshift-space distortions effects both in the linear and mildly non-linear regimes. The redshift-space auto- and cross-covariances for the multipoles of the anisotropic power spectrum and $2$-point correlation function are compared to the $N$-Body reference ones in Sec.~\ref{compzspace}.In Sec.~\ref{conclusion} we draw our conclusions.
\section{Theoretical pipeline in real space}
\label{pipelinecomovingspace}
In this section we recall the basics of the pipeline enabling the generation of a discrete set of objects defined by a given, arbitrary $1$-point PDF and density power spectrum. For a more detailed description of the method we refer the reader to \cite{Baratta:2019bta}.

In a first stage we focus on the generation of the continuous density field. Then we discuss how to circumvent the limit of the method, that is the production of negative, non-physical power spectra, when the target power spectrum is the non-linear one. Finally we describe how to discretise the field into point-like tracers.

\subsection{Targeting a given PDF and power spectrum}
\label{targeting}
The main ingredient of the pipeline lays on the local, non-linear mapping of an initial (centered and reduced) Gaussian field $\nu(\vec x) \equiv \mathcal{N}(0,1)$ such that 
\begin{equation}
\delta(x) \equiv \mathcal L \left[\nu(x)\right]\ ,
\label{mappingfunction}
\end{equation}
where the resulting $\delta$-field must represent the final cosmological, non-Gaussian contrast density field. The transformation $\mathcal L$ can be found using standard probability transformation rules, as long as a target $\delta$-PDF is set. In particular, using the local conservation of probabilities between $\nu$ and $\delta$, $\mathcal L$ in Eq.~\ref{mappingfunction} can directly be identified to
\begin{equation}
\delta = C_\delta^{-1}[C_\nu(\nu)]\ ,
\label{cumulativesforL}
\end{equation}
where $C_Y$ is defined as the cumulative function of the field $Y$. Eq.~\ref{mappingfunction} not only impacts the moments of $\nu$, but also its $2$-point statistics. Still using probability conservation properties, one can quantify the effect of applying a local transformation $\mathcal L$ to the Gaussian field through the mapping $\lambda$ by
\begin{equation}
\xi_\delta \equiv \left < \delta_1\delta_2 \right> = \iint \mathrm d\nu_1\mathrm d\nu_2\mathcal L(\nu_1)\mathcal L(\nu_2)\mathcal P_\nu(\nu_1,\nu_2,\xi_\nu)\equiv \lambda(\xi_\nu)\ ,
\label{relation2pstats}
\end{equation}
where $\mathcal P_\nu(\nu_1,\nu_2,\xi_\nu)$ is the bivariate normal distribution of $\nu$, and $\xi_\nu \equiv \left < \nu_1\nu_2 \right>_c $ its corresponding $2$-point correlation function. $\xi_\delta$ stands for the $2$-point correlation function of the cosmological field.

Since the power spectrum, $P(k)$, is the Fourier transform of the $2$-point correlation function, we can rely on the mapping $\lambda$ in order to predict the input power spectrum of the Gaussian field $\nu$. Indeed with $|\xi_\nu|<1$, $\lambda$ turns out to be continuous and monotonic in this interval. This implies the existence of the reciprocal function $\lambda^{-1}$, ensuring Eq.~\ref{relation2pstats} to be inverted as $\xi_\nu = \lambda^{-1}(\xi_\delta)$.

As a matter of fact, once $\xi_\delta$ is known (provided that $P_\delta(k)$ is the target power spectrum), being able to compute $\xi_\nu$ is equivalent to find out the power spectrum $P_\nu(k)$ to be assigned to the initial Gaussian field. This specific field is then generated such that, once transformed under the local mapping $\mathcal L$, the resulting $\delta$-field follows the targeted power spectrum $P_\delta(k)$.
Formally speaking, an overview of the pipeline reads

\begin{equation}
P_\nu(k) = \mathcal F \left\{   \lambda^{-1} \mathcal F^{-1}\left[ P_\delta(k) \right]   \right\}\ ,
\label{overalltransfo}
\end{equation}
where $\mathcal F$ and $\mathcal F^{-1}$ stand respectively for the Fourier and inverse Fourier transforms of the power spectra. 
The presented inversion scheme relies on our ability to evaluate the transformation $\lambda$ of the $2$-point correlation function. Therefore, we need to compute numerically Eq.~\ref{relation2pstats}\footnote{It can be shown that it is possible to switch the 2-D integral Eq.~\ref{relation2pstats} into a sum of finite 1-D integrals, quickly converging and therefore less computationally expensive. As a result, computing the Hermite expansion of the local mapping, $\mathcal L$, allows one to write down the Taylor expansion of the $2$-point correlation function $\xi_\nu$.}.

\subsection{Practical implementation : constrains and limits}
\label{practicalimplementation}
Let us consider a periodic, cubical comoving volume of size $L$. We assume that the different fields are sampled on a regularly spaced grid in this box, characterised by the sampling parameter $N_s$, representing the number of grid nodes per dimension. In this setting, we must now consider a three-dimensional description of the various fields and $2$-point statistics. In particular, the quantities described in Sec.~\ref{targeting} become $\nu(x) \rightarrow \nu(\vec x)$, $\delta(x) \rightarrow \delta(\vec x)$, $P(k) \rightarrow P(\vec k)$, etc.

As generating a non-Gaussian density field in configuration space requires to start from an initial Gaussian field in Fourier space\footnote{Generating a Gaussian field with a given power spectrum is greatly simplified in Fourier space.}, the Monte Carlo method extensively uses three-dimensional Fast Fourier Transforms ($\mathrm{FFT}$) and their inverse ($\mathrm{IFFT}$) in order to be efficient. As a result, we have to sample the generated fields on a regular grid in configuration space, producing in Fourier space a well known effect called aliasing \citep{Jing:2004fq}. 
Basically, the $\mathrm{FFT}$ of a sampled field is the standard Fourier transform of the field itself to which replicas (or aliases) of it are added. This produces an extra power on scales close to the Nyquist frequency $k_N = \pi/(L/N_s)$, also affecting the Fourier phases. Assuming that $\mathrm{IFFT}[\mathrm{FFT}[\vec Y]]=\vec Y$ (or equivalently that the $\mathrm{IFFT}$ of an aliased field is the "right" field in configuration space), the targeted power spectrum, $P_\delta(\vec k)$, used in Eq.~\ref{overalltransfo} must be artificially aliased according to
\begin{equation}
P_\delta^{\mathrm{aliased}}(\vec k) = \sum_{\vec n} P_\delta(|\vec k -2\vec n k_N|)\ ,
\label{aliasing}
\end{equation}
for the method to be self-consistent. Here $\vec n \equiv \left(n_1,n_2,n_3 \right) \in \mathbb Z^3$.
Simply stated, the set of fields we generate in Fourier space must necessarily be aliased. This ensures the fields in configuration space to be unaffected by aliasing.

Moreover, as the cosmological fields are sampled on a finite grid, this naturally introduces a smoothing scale of their statistics, related to the quantity $L/N_s$. As a consequence, prior to the application of Eq.~\ref{overalltransfo}, the two statistical inputs (the power spectrum and the PDF) must be smoothed consistently with the grid. This is natural for the PDF if its targeted shape is directly estimated on the same grid type, as well as for the power spectrum if it remains convolved with the mass assignment scheme needed for its estimation through a $\mathrm{FFT}$ \citep{sefusatti_15}.
However in the case of a target power spectrum which is free from any filtering effect (\textit{i.e.} if the effect of the mass assignment scheme has been removed from the estimated power spectrum or for a theoretical one), a scan filtering of the power spectrum must be considered in such a way that the variances of the two inputs of the method are matching, prior to the computation of its 3-D aliased version (following Eq.~\ref{aliasing}).
Such scan filtering can be performed using multiple filtering radius, $R_1$, values in 
\begin{align}
P^{filt}(k) = P(k)F(k) = P(k)\ \mathrm{exp} \left[ -(kR_1)^i \right] \,,
\label{filteringmodel1}
\end{align}
where $i$ is a free parameter weighting the filtering tail shape. For such a low pass filter, it should be noted that the aliasing contributions are significantly reduced. However, for the sake of accuracy, we still take them into account.

It is worth mentioning that this Monte Carlo procedure is not rigorously implementable to any shape of power spectrum, especially the one characterised by a non-linear prescription. They produce negative values in the 3-D power spectrum $P_\nu(\vec k)$ in Eq.~\ref{overalltransfo}, whether aliasing Eq.~\ref{aliasing} is taken into account or not. Such outcome is at this stage not understood. As an example, with a high resolution grid of $N_s=1024$ and $L=1000$ Mpc$/h$, corresponding to a precision of $\sim 1$ Mpc$/h$, about $\sim 45\%$ of negative elements are produced for a $\Lambda$CDM non-linear power spectrum at redshift $z=0$. They are mainly located in the Fourier volume in the vicinity of the Nyquist frequency.
This number of negative values is reduced for $N_s=512$ ($\sim 20\%$) and disappears for $N_s=256$. But choosing such poor grid setting would inevitably reduce the range of usable simulated modes.

Although, on the one hand the variance $\sigma_\nu^2 = k_F^3\sum_i P_\nu (\vec k_i)$, estimated from the power spectrum of the Gaussian field ($k_F=2\pi/L$ being the fundamental frequency of the periodic box), is actually well unitary as required by the procedure, on the other hand these non-physical negative elements prevent the simulation of the Gaussian field $\nu_{\vec k}\sim \sqrt{P(\vec k)}$.

To sidestep this restriction, a corrective method can be adopted: it consists in clipping to zero all negative values. 
As a direct consequence, the variance computed on the new 3-D power spectra automatically deviates from unity with $\sigma_\nu^2 >1$.
In order to recover the procedure condition, $\sigma_\nu^2 =1$, an additional 3-D filtering can be applied on it to match the unit variance. Here we propose the 3-D filtering version of Eq.~\ref{filteringmodel1}, \textit{i.e.}
\begin{equation}
    F(\vec k) = \mathrm{exp} \left[ -(\vec kR_2)^j \right]\ .
    \label{filteringmodel2}
\end{equation}
Naturally, this procedure has the effect of reducing the maximum mode up to which the power spectrum is reliably simulated. While $R_1$ and $R_2$ are constrained by the previously discussed conditions on the variance, $i$ and $j$ should be chosen in such a way that the simulated power spectrum matches the targeted one up to a maximum wave-mode. Their values are discussed in Sec.~\ref{resultscomovingspace}.

Let us briefly summarise the method to provide the reader with a global understanding of the method. We generate in Fourier space a Gaussian field, $\nu$, on a 3-D grid defined by the power spectrum $P_\nu(\vec k)$ (Eq.~\ref{overalltransfo}), after being corrected using Eq.~\ref{filteringmodel2}. We then perform an inverse $\mathrm{FFT}$ of $\nu(\vec k)$ to get $\nu(\vec x)$ and locally apply on it the non-linear transform, $\mathcal L$, in Eq.~\eqref{mappingfunction}. The obtained field, $\delta(\vec x)$, is now expected to both follow the target density PDF (the exact one) and (filtered) power spectrum.

\subsection{From continuous field to discrete catalogues}

In order to finalise the procedure above, one can turn the obtained density field, $\delta(\vec x)$, into a discrete catalogue of objects by applying a local Poisson sampling \citep{1956AJ.....61..383L} of it. Provided a mean number density, $\rho_0$, expected in a catalogue of volume $V$, and a simulated local density contrast field, $\delta(\vec x)$, one can estimate the number of objects, $\Lambda$, locally expected within a small volume element, $a^3=\left(L/N_s\right)^3$, as $\Lambda = \rho_0 a^3 (1+\delta)$.

Of course, this expected number $\Lambda$ is not an integer. However it can be seen as the expectation value that one should see if the sampling was made several times at position $\vec x$. Thus the number of observed objects in the small volume $a^3$ can be drawn from a Poisson distribution with expectation value $\Lambda$. 

One can think about two kinds of interpolation schemes when placing the corresponding number of particles within the grid cell. The first is the so-called Top Hat scheme, where particle coordinates follow a uniform distribution. This is fast but the density fields will appears structurally discontinuous. A natural extension of this method, able to fix this issue, is the three-linear scheme, implemented here by populating the grid cell with particles following the three-linearly interpolated density field distribution.

Note that both methods have an impact on the shape of the resulting Poisson power spectrum, that fortunately can be accurately predicted even beyond the Nyquist frequency \citep[see][]{Baratta:2019bta}.

To anticipate our outcomes, we find that the choice of the interpolation scheme has no impact on the estimated covariance matrix in the considered wave-mode range. Still in the following, we only use the three-linear interpolation scheme.
\section{Application in real space}
\label{resultscomovingspace}
The previously introduced method allows one to simulate within a cubical volume objects whose nature (dark matter particles, dark matter halos, galaxies) only depends on the choice of the input density power spectrum and PDF.
In this section, for simplicity, we choose to simulate dark matter particles in real space when targeting the statistics of a set of $N$-body simulations. We evaluate the range of reliability of the produced $2$-point statistics for a reasonable (computationally speaking) grid sampling precision of $N_s=1024$.

\subsection{The \texttt{DEMNUni\_cov} $N$-body simulations}
Originally developed for testing different probes in the presence of massive neutrinos and dynamical dark energy, the \texttt{DEMNUni} experiment is a suit of $N$-Body simulations \citep{carbone_16, castorina_15, parimbelli_22} in $\Lambda$CDM plus different $\nu w_0w_a$CDM cosmological scenarios. The largest simulations follow the evolution of 2048$^3$ CDM and (if present) 2048$^3$ neutrino particles, in a box of side $L=2 \ \mathrm{Gpc}/h$. They have been produced on various High Performance Computing (HPC) machines provided by the super computing centre CINECA\footnote{\href{http://www.cineca.it/}{cineca.it}}, Italy, by exploiting the GADGET-3 code of~\citet{springel_05}, further developed by~\citet{viel_10} for the inclusion of massive neutrino particles.

All the simulations of the \texttt{DEMNUni} have in common the following cosmological parameters:
\begin{eqnarray}
\Omega_k & = & 0\ , \nonumber \\
\Omega_m & = & 0.32\ , \nonumber \\
\Omega_b & = & 0.05\ , \nonumber \\
h & = & 0.67\ , \nonumber \\
n_s & = & 0.96\ , \nonumber \\
A_s & = & 2.1265\times 10^{-9}\ . \nonumber
\end{eqnarray}
The initial conditions have been set at redshift $z=99$ using the Zeldovich approximation \citep{zeldovich_69} and the public code described in~\citet{zennaro_16},~\citet{zennaro_17} and~\citet{zennaro_19}.

The evolution of cosmic structures is simulated down to redshift $z=0$, while outputting $63$ particle snapshots with equally spaced logarithmic scale factor intervals. Among these snapshots, we restrict the analysis of this work to the five following redshifts\footnote{For simplification in the following, these five redshifts will be referred to the round numbers $z=\{ 0,0.5,1,1.5,2\}$. }. 
$$z=\{ 0,\; 0.48551,\; 1.05352,\; 1.45825,\; 2.05053\}.$$
Regarding our study oriented toward the estimation of covariances, only a \texttt{DEMNUni} subset, hereafter called \texttt{DEMNUni\_cov}~\cite{parimbelli_21}, will constitute the reference basis for the validation tests of our Monte-Carlo method. This subset is made of two $\Lambda$CDM cosmologies, one without massive neutrinos and the other one with massive neutrinos (added at the level of massive particles inside the simulation) of total mass $M_\nu=0.16$ eV. In the following these two cosmologies will be referred as '$\Lambda$CDM' and '$16nu$', respectively. The main interest of these two suits of simulations is that they have been ran $50$ times each with different realisations of the initial conditions. This way they offer the possibility of estimating to some extent covariance matrices. 

Each simulation consists in $1024^3$ dark matter particles of mass $ m_p\simeq 8\times 10^{10} M_\odot/h$, and $1024^3$ massive neutrino particles (in the case of the massive neutrino cosmology) of mass $m_p^\nu\simeq 1\times 10^{9} M_\odot/h$ in a cubical volume of size $L = 1000$ Mpc$/h$ with periodic boundary conditions. 
For each snapshot of each realisation, we estimate the PDF and the power spectrum of the matter field using a Peace-wise Cubic Spline (PCS) mass assignment scheme on a grid of sampling parameter $N_s=1024$. The power spectrum is estimated using the \texttt{NBodyKit}\footnote{\href{https://nbodykit.readthedocs.io/en/latest/}{nbodykit.readthedocs.io}} \citep{hand_17} software, applying the interlacing method to reduce the aliasing \citep{sefusatti_15}.

The averaged PDF and filtered power spectra (following prescription in Eq.~\ref{filteringmodel1}) of the $50$ simulations can then be directly used as targets for the Monte Carlo code.

\subsection{Comparing the $2$-point statistics}

\begin{figure}
\begin{center}
\hspace*{-0.3cm}\vspace*{-0.0cm}\includegraphics[scale=0.52]{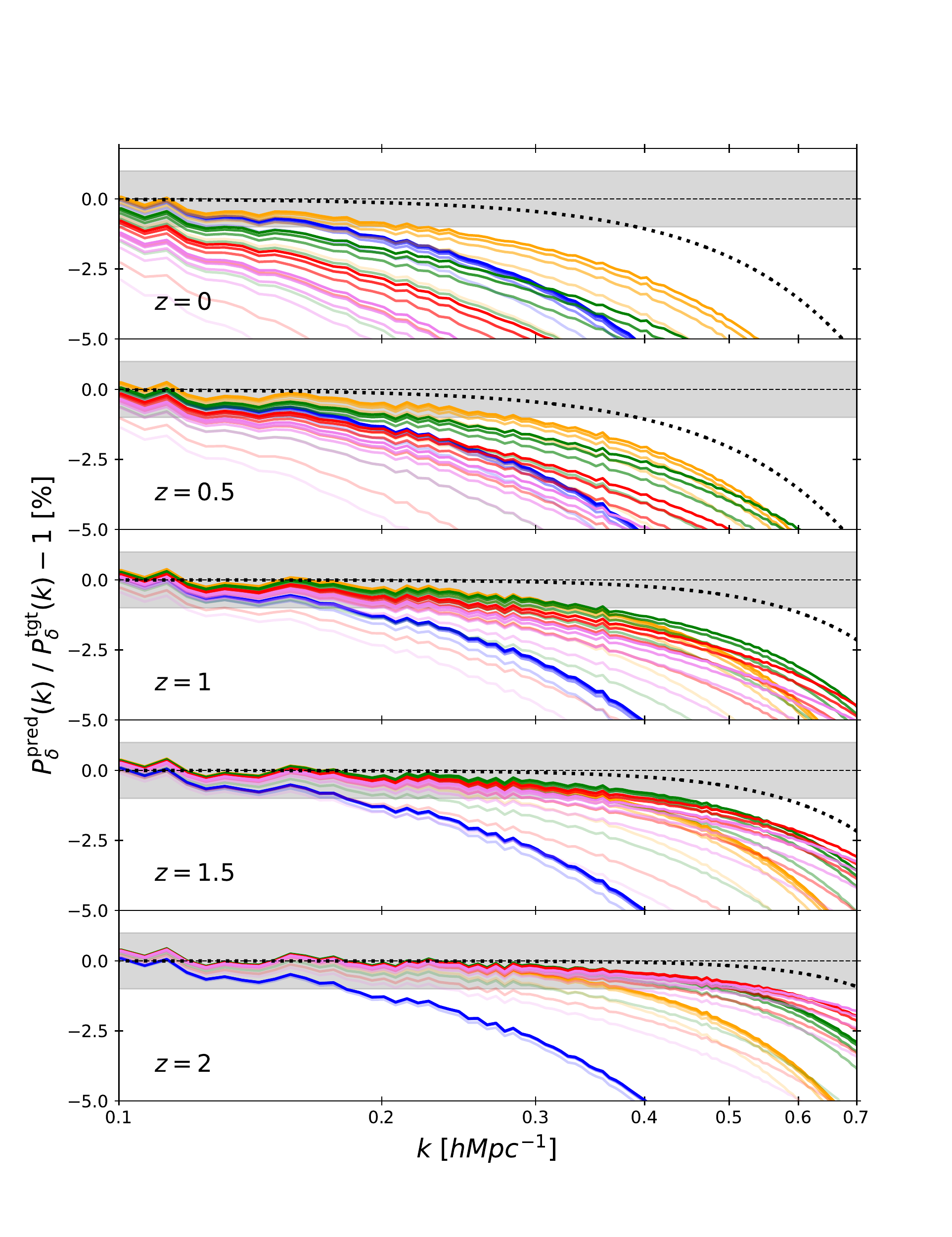}
\end{center}
\caption{Relative deviation between the predicted grid power spectrum (after the two filterings Eq.~\ref{filteringmodel1} and~\ref{filteringmodel2}) and the corresponding target power spectrum for all $5$ selected redshifts in the $\Lambda$CDM cosmology (similar results are obtained for the massive neutrinos, $16nu$ cosmology). The predicted spectra are computed for all combinations of parameters $i=[2,3,4,5,6]$ and $j=[2,4,6,8,10]$. Each colour represents a certain value of $i$, respectively blue, orange, green, red and violet. An increasing colour intensity means an increasing $j$ value. The dotted black line is the relative deviation when only the first filter, Eq.~\eqref{filteringmodel1}, is applied, taking the identified optimal $i$, respectively $3, 3, 4, 4$, and $5$. The grey area delimits the $1\%$ limit.}
\label{fig:optimali1i2}
\end{figure}

In the first place, for the Monte Carlo simulations to be run we must fix the values of the filtering parameters $i$ and $j$ (see eqs.~\ref{filteringmodel1} and~\ref{filteringmodel2}). For various combinations of them, with $i$ taking several intermediate values in $[2,6]$ and $j$ in $[2,10]$, one can predict the output $P_\delta(\vec k)$ of the method by explicitly computing $\texttt{FFT}[\lambda(\xi_\nu)]$ (see Eq.~\ref{relation2pstats}). We can shell average the resulting 3-D power spectra in shells of width $k_F$ and compare them to the targeted one. The results, presented in Fig.~\ref{fig:optimali1i2}, show that, in order to maximise the range of well simulated modes, $i$ must be equal to $3$ for redshifts $z=[0,0.5]$, then $i=4$ for $z=[1,1.5]$ and finally $i=5$ for $z=2$. On the other hand, an arbitrarily large value for j can be chosen. Fig.~\ref{fig:optimali1i2} only presents the results for the $\Lambda$CDM scenario. Note that in the present case as well as for the following figures, only one cosmology will be discussed if the same results are obtained also for the other one.

Moreover, from this figure, we can see the successive effects of the two filtering, Eq.~\eqref{filteringmodel1} and \eqref{filteringmodel2}. The difference between the dotted black line and zero is due to the first filter. We recall that this smoothing is exclusively due to the fact that the fields are set on a finite grid. The difference between the dotted black line and the coloured ones shows the impact of the second filter, required to solve issue of the negative elements of the spectrum.

For the present analysis, in the following we use the same doublet parameters $(i,j)$ for all redshifts and cosmologies, fixing them to $(i=2.5,j=8)$. This choice is first motivated by the fact that we want to carry out a comparative study of the method between the redshifts, without adding extra tuning to enhance the output statistics. Second, we recall that the PDF is obtained from a PCS mass assignment scheme. This has the particularity of acting as a filter on the power spectrum fairly well reproduced by Eq.~\eqref{filteringmodel1} with $i=2.5$. With such a choice, we try to keep a good consistency between these two statistics and prevent eventual pollution on the covariances.

We ran $5,000$ simulations of \texttt{COVMOS} catalogues per redshift in the $\Lambda$CDM and $16nu$ cosmologies. Each of them consists in $10^8$ dark matter particles in boxes of size $L=1000$ Mpc$/h$. We recall that the target statistics of the method are the ones estimated on the \texttt{DEMNUni\_cov} simulations. 

\begin{figure}
\hspace*{-0.4cm}\vspace*{-0.0cm}\includegraphics[scale=0.515]{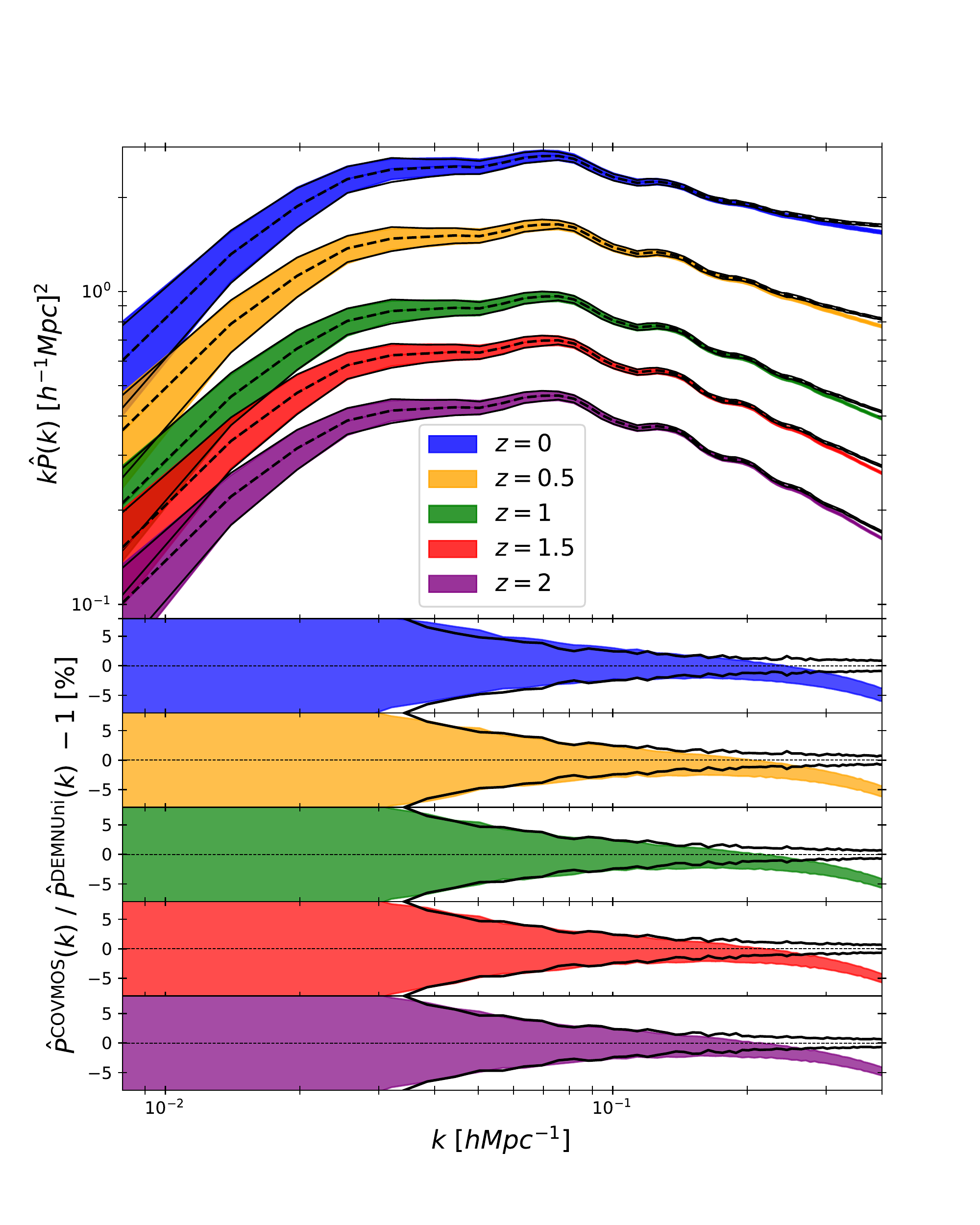}
\caption{\textit{Top panel}: estimated and averaged real space power spectrum monopoles over $5,000$ realisations of \texttt{COVMOS} catalogues in $16nu$ cosmology, with dispersion on each realisation represented in colour. The same quantities are represented in black for the $50$ \texttt{DEMNUni\_cov} realisations. \textit{Bottom panels}: The relative deviations between the averaged \texttt{COVMOS} and \texttt{DEMNUni\_cov} measurements with error bars.}
\label{fig:pkcomov}
\end{figure}

After removing the relatively low shot noise contribution to the estimated power spectra ($P_{SN}\sim 0.04h^{-3}$Mpc$^3$), we compare them to the ones estimated from the \texttt{DEMNUni\_cov} in Fig.~\ref{fig:pkcomov}. As expected, we can observe a smaller range of reliable simulated wave modes when compared to its grid equivalent in Fig.~\ref{fig:optimali1i2}. Indeed, as already discussed, the Poisson sampling gives rise to a smoothing of the cosmic field that visually starts to dominate around $k\sim0.2~h/$Mpc. This smoothing has the direct consequence to reduce to around $k \sim 0.25h/$Mpc (in the $1$-$\sigma$ limit) the maximum accurately simulated Fourier mode. We recall that this result is obtained assuming a grid precision of $N_s=1024$. Improving this setting would automatically increase the range of well simulated modes, at the cost of a longer running time.
It is also noted that this filtering effect is similar, regardless of the redshift. This is due to the fact that we have defined the parameters $i$ and $j$ as constants.

\begin{figure}
\hspace*{-0.7cm}\vspace*{-0.0cm}\includegraphics[scale=0.42]{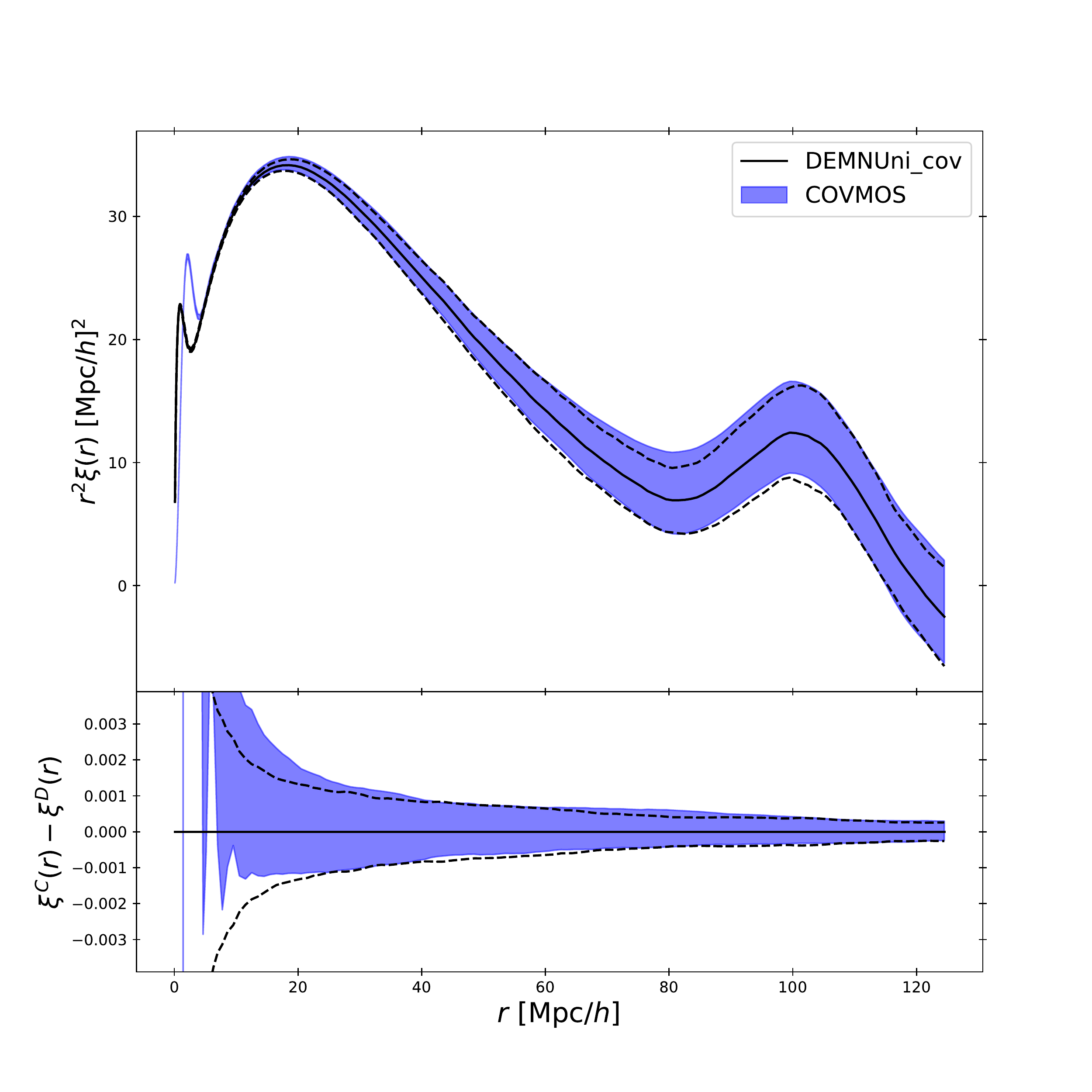}
\caption{\textit{Top}: estimated and averaged correlation function monopoles over $100$ realisations of \texttt{COVMOS} catalogues in $16nu$ cosmology at $z=0$ with dispersion on each realisation represented in blue. The solid black line represents the averaged power spectra (with dispersion over measurements in dashed black lines) over 50 \texttt{DEMNUni\_cov} simulations. \textit{Bottom}: The residual between the \texttt{COVMOS} and \texttt{DEMNUni\_cov} outputs with error bars.}
\label{fig:xicomov}
\end{figure}

When switching to the configuration space, we estimated the $2$-points correlation functions over the $50$ \texttt{DEMNUni\_cov} realisations at $z=0$ and for the $16nu$ cosmology only. The outputs are compared to a set of $100$ correlation functions estimated on the \texttt{COVMOS} set of simulations. We restrict this number given the high computational cost of estimating such statistics. Fig.~\ref{fig:xicomov} shows this comparison in the range of scales $[0.1,125]$ Mpc$/h$, depicting a high level of agreement between the two methods for all scales higher than $7$ Mpc$/h$.
Below this limit, correlations seem over-produced in the range $[1.5,5]$  Mpc$/h$ and under-produced for scales $<1.5 $  Mpc$/h$.

\section{Theoretical pipeline in redshift-space}
\label{pipelineredshiftspace}

In galaxy surveys the observed positions of tracers of the matter field are systematically affected by their peculiar velocities. 
As a consequence, the power spectrum is affected at all scales, this is the so-called redshift-space distortion (RSD) effect. Indeed, it acquires an angular dependence with respect to the line-of-sight, loosing its property of being isotropic. The amplitude of its angular average is enhanced in the linear regime due to the coherent in-fall of objects toward high density regions, this is the so-called Kaiser effect \citep{kaiser_87}. In the non-linear regime (on smaller scales) the density fluctuations and velocities are coupled in a non trivial way \citep[][]{Scoccimarro:2004tg} which make the RSD effects on the power spectrum difficult to be predicted accurately.

In the present section we show our practical implementation of peculiar velocities in the \texttt{COVMOS} code.
The main idea is to first generate a velocity field correlated with the density field so that we make sure that we are able to reproduce the Kaiser effect on large scale.

In principle, applying the Monte Carlo pipeline as detailed in section \ref{pipelinecomovingspace} to generate a continuous velocity field (as long as a corresponding power spectrum and a PDF are provided) is perfectly feasible. However, several subtle points must be addressed. 

The first is related to the fact that we expect the divergence of the peculiar velocity field $\vec v_p$ and the density field $\delta$ to be correlated to satisfy the linear continuity equation 

\begin{equation}
    \nabla \cdot \vec v_p = -a H(a)f \delta(\vec x), 
\label{continuity}
\end{equation}
where $f$ is the growth rate of structures.
This implies that the velocity field cannot be simulated in an independent way from the density field $\delta$. 

Beside, once the velocity field has been generated on a regular grid in space, one needs to carefully associate a velocity to the previously generated tracers anywhere in the simulated volume. In the following we detail these two main steps in order to simulate a catalogue of objects together with their peculiar velocities.

\subsection{Simulating the velocity field}

The method that we developed to generate the velocity field is based on the fact that when estimating the velocity field with a Delaunay tesselation in the DEMNUni simulations \citep[as in][]{bel_18} the PDF of each velocity component appears to be close to be Gaussian (see the black dashed line in Fig.~\ref{fig:PDFs_velo} as compared to the Gaussian model in red dashed). In addition, we want to reproduce the power spectrum of the divergence of the velocity field  and finally, as stated before, we need to satisfy the continuity equation.

Let us define the scaled velocity field $\vec u (\vec x)$ as
\begin{align}
 \vec u (\vec x) \equiv -\vec v_p(\vec x) \frac{1+z}{100E(z)},
 \label{ux}
\end{align}
where $E(z) \equiv H(z)/H_0$, such that the observed comoving distance $s$ of an object can be deduced from the true one $r$ as $s = r - u_r$ ($u_r$ being the line of sight projection of $\vec u$). The idea, is then to generate the divergence $\theta (\vec x) = \vec \nabla \cdot \vec u$ of the scaled velocity field in Fourier space according to the power spectrum of the divergence of the velocity field and to assume that the velocity field $\vec u$ is curl free 
\begin{equation}
\vec{u}_{\vec k} = -i \frac{\vec k}{k^2}\theta_{\vec k}\ .
\label{inversescalar}
\end{equation}
This way it is only necessary to generate the divergence $\theta_{\vec k}$ in Fourier space with a real and an imaginary part following a centred Gaussian distribution with a variance given by the power spectrum of the divergence $P_{\theta\theta}(k)$. Notice, however, that in order to satisfy to the continuity equation we impose that the Gaussian field $\nu$ used to generate the density field is also used to generate the divergence. This means that we impose strictly the continuity equation only in the linear regime.
To do so one can rely on the fitting functions given in~\citet{bel_18}
\begin{align}
P^\mathrm{lin}_{\theta\theta}(k) &= f^2(k) P^\mathrm{lin}_\mathrm{cb}(k) \ ,\label{Jufitfunclin}\\ 
P_{\theta\theta}(k)  &= P^\mathrm{lin}_{\theta\theta}(k)\ \mathrm{exp}\left(-a_1k-a_2k^2-a_3k^3\right)\ , \label{Jufitfunc}
\end{align}
where $P^\mathrm{lin}_\mathrm{cb}(k)$ stands for the linear cold dark matter plus baryons power spectrum (numerically predicted by Boltzmann codes) and
\begin{align}
a_1 &= -0.817 + 3.198\sigma_{8,m}\ \\
a_2 &=  +0.877 - 4.191\sigma_{8,m}\ \\
a_3 &= -1.199 + 4.629\sigma_{8,m}\ ,
\end{align}
where $\sigma_{8,m}$ is the $rms$ of the linear matter fluctuations smoothed on spheres of radius $8\ \mathrm{Mpc}/h$ and must be predicted for the total matter (accounting for massive neutrinos).

In summary, one first simulate on a Fourier grid the Gaussian $\theta$-field using the same random Fourier phases as for the density field, defined by the non-linear $\theta-\theta$ power spectrum in Eq.~\ref{Jufitfunc}. This field is then converted in the $3$-dimensional velocity field in Eq.~\ref{inversescalar}, on which we apply an inverse Fourier transform to obtain the velocity field in configuration space, $\vec u (\vec x)$, of~Eq.\eqref{ux}.

\subsection{Drawing peculiar velocities}

In order to properly reproduce redshift space distortions, it is necessary to assign velocities to the objects inside the simulated catalogue. To this aim we could interpolate the regularly spaced velocity grid $\vec u (\vec x)$ at all the positions of the objects, however this would not accurately reproduce the non-linear redshift space distortions. Indeed, on small scale one expect that the angular average power spectrum is suppressed by the fact that inside haloes velocities are not coherent but they are close to be random thanks to the virial theorem. That is the reason why we draw the velocity of each object out of a Gaussian distribution centred on the interpolated value of the velocity field $\vec u$ at the position of the object and with a given dispersion $\Sigma$. This is a similar method to the one used by~\citet{delatorre2013} in order to assign velocities to haloes below the mass resolution of a simulation.  

\begin{figure}[ht!]
\begin{center}
\hspace*{-0.35cm}\vspace*{-0.0cm}\includegraphics[scale=0.68]{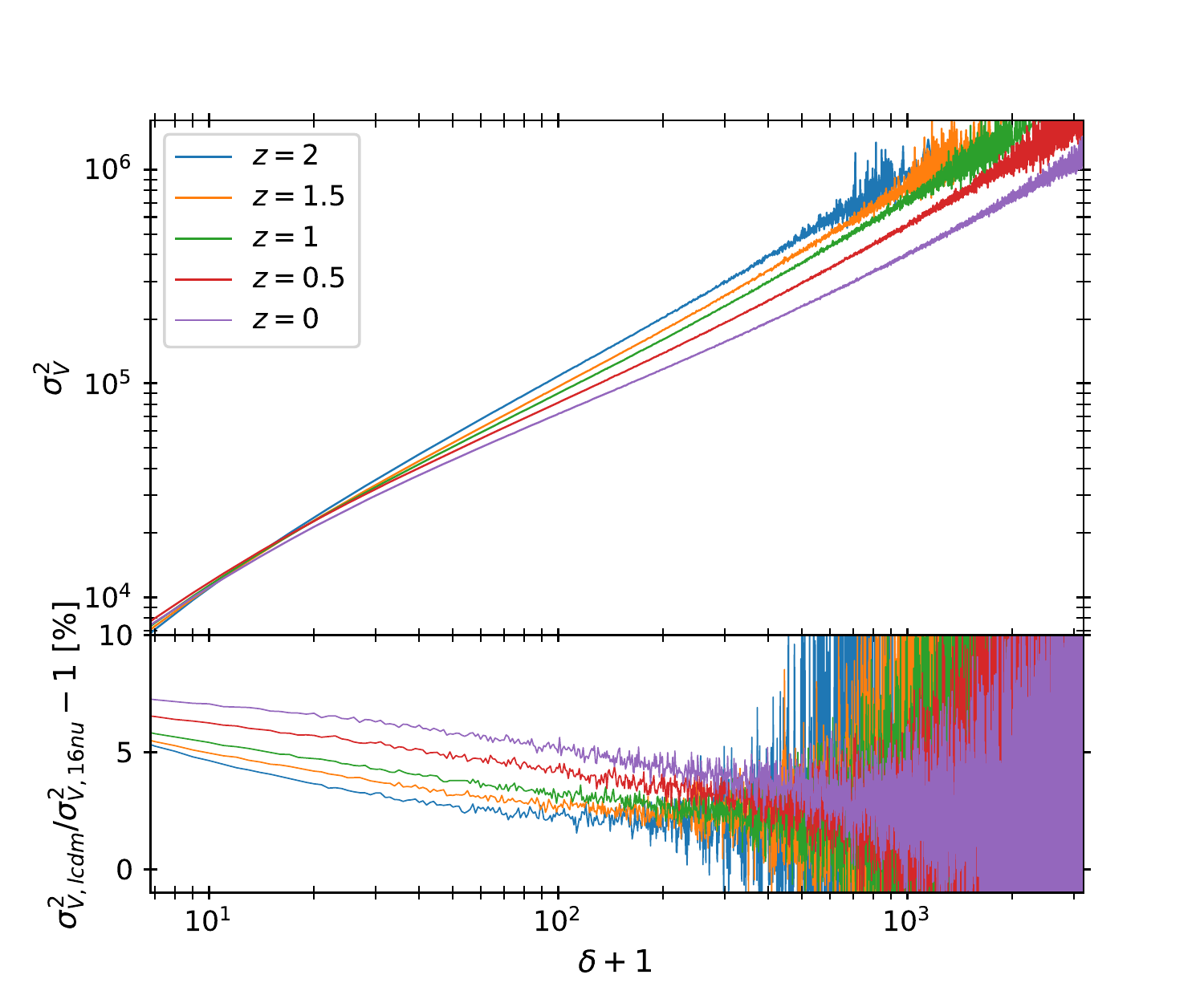}
\end{center}
\caption{\small \textit{Top}: Combined velocity components variance in (km/s)$^2$ as a function of local density contrast for the \texttt{DEMNUni\_cov} set of 50 simulations and for the five redshifts $z=[0,0.5,1,1.5,2]$ in the case of the standard $\Lambda$CDM cosmology. \textit{Bottom}: Percent relative deviation, redshift per redshift, between the two cosmologies, $\Lambda$CDM and 16nu, still using the same colour-code.}
\label{fig:variancesVSdeltaDEMNUni}
\end{figure}

Aiming at being as realistic as possible, we allow the velocity dispersion to depend on the underlying matter density. In fact, one expects that more massive haloes have a large velocity dispersion \citep{Elahi:2017cjx}. Using the \texttt{DEMNUni\_cov} simulations we estimated the relation between the local density $1+\delta$ and the velocity variance $\sigma_V^2$. By fixing the grid sampling to $N_s=1024$, one can run an estimation of the local velocity variance as a function of the local density contrast value over the $50$ simulations in the two cosmologies $\Lambda$CDM and $16nu$, for the five redshifts $z=\left[0,0.5,1,1.5,2 \right]$. This is represented in Fig.~\ref{fig:variancesVSdeltaDEMNUni} where, in the upper panel one can see that, no matter what is the considered redshift, there is a clear relation between the velocity dispersion and the density. Indeed, this relation exhibits a very low level of stochasticity meaning that we can further assume that there is an analytical relation between the local dispersion and the density. In addition, the lower panel of Fig.~\ref{fig:variancesVSdeltaDEMNUni} shows that the presence of massive neutrinos induces changes in this relationship. As expected, the dispersion in a cosmology with massless neutrinos is systematically higher. 

Inspecting Fig.~\ref{fig:variancesVSdeltaDEMNUni} one can see that the global behaviour of the velocity dispersion-density relation can be approximated as a power law. 
Even if this does not seem to be accurate for low values of the density we will further assume a power law relationship. This is justified for two reasons:
\begin{itemize}
    \item When the density is low, so is the dispersion. Thus the way we model the relation is expected to affect less the final velocities of the various objects in the catalogues;
    \item When the density is low, we estimate both the density and the velocity dispersion with a low number of objects, thus the shot noise can introduce a bias.
\end{itemize}
Thus in the end we adopt the following analytical form 
\begin{equation}
\Sigma^2(\rho) \equiv \beta \rho^\alpha , 
\label{modelforsigma}
\end{equation}
where $\alpha$ is fixed from the velocity dispersion-density relation estimated from \texttt{DEMNUni\_cov} and listed in the right column of table \ref{tableonepvd}. The second parameter $\beta$ is not taken from the simulations, but it needs to be set in order to make sure that the $1$-point velocity dispersion $\sigma_V$ of the catalogues is matching the one of \texttt{DEMNUni\_cov} (this will be explained more in details in the following).

Note also that the exponent $\alpha$ of the power law can be related to dynamics inside over-densities, thus by extension inside haloes. In fact, we checked that if one uses haloes to estimate the velocity dispersion-density relation, then the same value for $\alpha$ is measured but with a different normalisation $\beta$.
\begin{table}
\centering 
\begin{tabular}{l|c c|c c}
    
           & $\sigma_V$($\Lambda$CDM) & $\sigma_V$($16nu$)  & $\alpha$($\Lambda$CDM) & $\alpha$($16nu$)  \\ \hline
$z=0$          &   $3.9412$    &  $3.7870$& $ 0.7437$ & $0.7544$\\  \hline
$z=0.5$        &   $3.6393$   &   $3.5032$& $ 0.8277 $ & $0.8368$\\ \hline
$z=1$           &   $3.1772$   &   $3.0640$& $ 0.9032$ & $0.9086$\\ \hline
$z=1.5$        &   $2.8930$   &   $2.7936$ &  $ 0.9363$ & $0.9414$\\ \hline
$z=2$           &  $2.5650$    &   $2.4818$  & $ 0.9594$ & $0.9554$\\ 
     
\end{tabular}
\caption{Averaged one-point velocity $rms$ (in Mpc/$h$) of the one-point velocity distribution as measured from particles from the 50 \texttt{DEMNUni\_cov} simulations, per redshift and per cosmology. The $\alpha$ parameters are inferred from the same set of simulations using linear regressions in ranges $\delta +1 \in [50,500]$.} 
\label{tableonepvd}
 \end{table}

The measured $1$-point velocity variances are listed in the left column of table \ref{tableonepvd}, where we can see that it decreases with the redshift and that adding massive neutrinos is reducing it. As anticipated, we fix the normalisation $\beta$ so that the one-point velocity variance $\sigma_V^2$ is the same as the $1$-point velocity variance measured in the \texttt{DEMNUni\_cov} catalogues. 

Furthermore, matching the $1$-point velocity variance allows us to provide a physical description of the amplitude parameter $\beta$. Thus, we can show how the $1$-point variance is related to the prescription in Eq.~\eqref{modelforsigma} when going from the large scale velocity to the velocity of individual objects. Indeed, the peculiar velocity, $\vec V_p$, assigned to a particle, can be formally expressed as 
\begin{equation}
    \vec V_p = \vec w + \Sigma(\rho)\vec v,
    \label{vp}
\end{equation}
where $\vec w = -aH \vec u$ and $\vec v$ is drawn from an  uncorrelated tri-variate centred reduced Gaussian distribution. This means that the conditional probability distribution, $\pcal (V_c| w_c, \rho)$, of a single component, $V_c$, of the vector $\vec V_p$ is given by 
\begin{equation}
\pcal (V_c| w_c, \rho) = \frac{1}{\sqrt{2\pi}\Sigma(\rho)}e^{-\frac{(V_c-w_c)^2}{2\Sigma^2(\rho)}}.    
\label{PDFdispvsdelta}
\end{equation}
Since we define the variance of a single velocity component as the arithmetic mean of the square of the velocity component, it turns out that 
\begin{equation}
\sigma_V^2 = \left\langle \rho V_c^2 \right\rangle = \iiint \mathrm dV_c \mathrm dw_c \mathrm d\rho\ \rho V_c^2 \mathcal P(V_c,w_c,\rho), 
\label{sigmav_start}
\end{equation}
due to the fact that the density of points is not uniformly distributed, thus regions with higher density will provide a larger contribution to the expectation value. 

From Eq.~\eqref{sigmav_start}, in order to predict the $1$-point velocity variance $\sigma_V^2$, one needs to express the $1$-point probability distribution of the object velocity component, $V_c$, which depends on the density and the velocity, $w_c$, as
\begin{equation}
\mathcal P(V_c,w_c,\rho) = \pcal(V_c | w_c, \rho) \mathcal P(w_c,\rho). 
\label{spliting_1pvelo_Pr_distrib}
\end{equation}

Substituting Eqs.~\eqref{PDFdispvsdelta} and \eqref{spliting_1pvelo_Pr_distrib} in expression Eq.~\eqref{sigmav_start}, and using the fact that the second order moment of the Gaussian distribution reduces to the sum of its variance and the squared of its expectation value, Eq.~\ref{sigmav_start} simplifies in
\begin{equation}
\sigma_V^2 = \iint \mathrm dw_c \mathrm d\rho\ \mathcal P(w_c,\rho)\rho\left[  w_c^2  + \Sigma^2(\rho)\right] = \sigma_w^2 + \sigma^2_\Sigma\ .
\label{generalvariance}
\end{equation}
In the above expression the first term, $\sigma_w^2 \equiv \int \mathrm dw_c \mathrm d\rho\ \rho w_c^2 \mathcal P(w_c,\rho)$, is the variance of the interpolated velocity field that can be estimated independently from the choice of the function $\Sigma^2(\rho)$, while this is not the case for the second marginalised term, defined as $\sigma^2_\Sigma \equiv \int \mathrm d\rho \rho \Sigma^2(\rho)$.
Then coming back to our simple model in Eq.~\ref{modelforsigma}, the $\beta$ parameter can be obtained using Eq.~\ref{generalvariance} as
\begin{equation}
\beta = \left(\sigma_V^2  - \sigma_{w}^2\right)/\sigma_\alpha^2\ ,
\label{beta_pred}
\end{equation}
where $\sigma_\alpha^2 \equiv \left<  \rho^{\alpha+1} \right> = \frac{1}{N}\sum_i \rho_i^{\alpha}$ is directly estimated on the interpolated density field at the object coordinates. 
\begin{figure}
\hspace*{-0.4cm}\vspace*{-0.0cm}\includegraphics[scale=0.51]{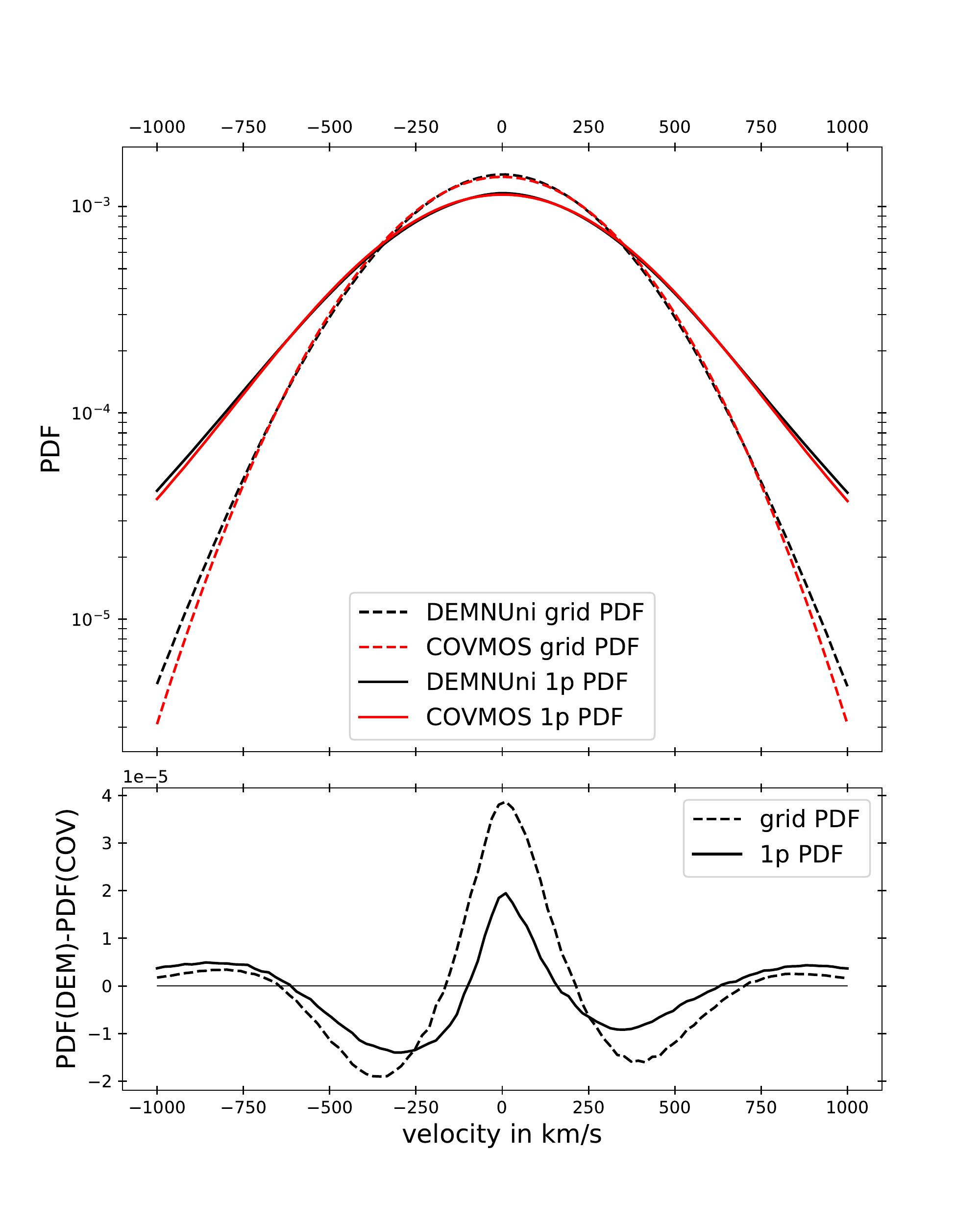}
\caption{\textit{Top}: Estimated one-point (solid lines) and smoothed (dashed lines) velocity PDF from one \texttt{COVMOS} simulation in red and one \texttt{DEMNUni\_cov} simulation in black. The smoothed velocity PDF is estimated from a Delaunay tesselation of the particle coordinates of the \texttt{DEMNUni\_cov\_01} experiment. This map is then interpolated on a grid of sampling parameter $N_s=512$. Here the cosmology is $\Lambda$CDM. \textit{Bottom}: residual between the \texttt{DEMNUni\_cov} and \texttt{COVMOS} PDF.}
\label{fig:PDFs_velo}
\end{figure}

In Fig.~\ref{fig:PDFs_velo} we compare the distribution of the velocity components of the regularly sampled velocity field $\vec w$, estimated from the \texttt{DEMNUni\_cov}, with the same distribution obtained from \texttt{COVMOS}. We see that the Gaussian approximation assumed to generate the velocity field is indeed a good approximation. In addition, we compare the $1$-point velocity PDF estimated from a \texttt{COVMOS} realisation of (exceptionally) $\sim 10^9$ dark matter particles with the one estimated from particles in a \texttt{DEMNUni\_cov} realisation. One can see that, assuming a power law relation between the velocity dispersion and the matter density (Eq.~\ref{modelforsigma}), and fixing the normalisation of the relation by requiring that the $1$-point velocity dispersion of \texttt{COVMOS} matches the one of \texttt{DEMNUni\_cov} (Eq.~\ref{beta_pred}), allow us to reproduce the non-Gaussian distribution of the particle velocity. Indeed, by matching only the second order cumulant moment, we also reproduce the exponential tails of the distribution.

In the next section we show the efficiency of our approximations for generating the velocity field at the level of the power spectrum in redshift-space. 
\section{Application in redshift-space}
\label{compzspace}
In this section we apply the whole pipeline for simulating particles (described in section \ref{pipelinecomovingspace}) and their associated peculiar velocities (section \ref{pipelineredshiftspace}). We study to which extent the various $2$-points statistics, associated error bars and correlations are reliably simulated.

\subsection{Validating the $2$-point statistics}
With the same settings as in section \ref{resultscomovingspace}, we simulated $5,000$ catalogues of $10^8$ dark matter particles with associated peculiar velocities, for all $5$ redshifts and for the two cosmologies. We estimate on each realisation the multipoles of the power spectrum 
\begin{equation}
P^{(\ell)}(k) = \frac{2\ell+1}{2}\int_{-1}^1 \mathrm d\mu_k P( k,\mu_k) \lcal_\ell(\mu_k)
\label{multipolepowerspectrum}
\end{equation}
where $\lcal_\ell(x)$ are defined as the Legendre polynomials of order $\ell$, $\mu_k \equiv \cos(\phi_{\vec k})$ and $\phi_{\vec k}$ is the angle between the line-of-sight and the wave vector $\vec k$. We adopt the plane parallel limit which consists in choosing the line-of-sight as being a fixed direction in the simulated comoving volume.

Generating the entire set of simulations and measuring their associated real and redshift-space power spectra took nearly $20$ days, running on about $940$ processors of $2.4$ GHz. 
Note that reducing the precision of the \texttt{COVMOS} grids to $N_s=512$, $3$ days are necessary to simulate and estimate the power spectrum of $100,000$ catalogues.

\begin{figure}
\hspace*{-0.55cm}\vspace*{-0.0cm}\includegraphics[scale=0.51]{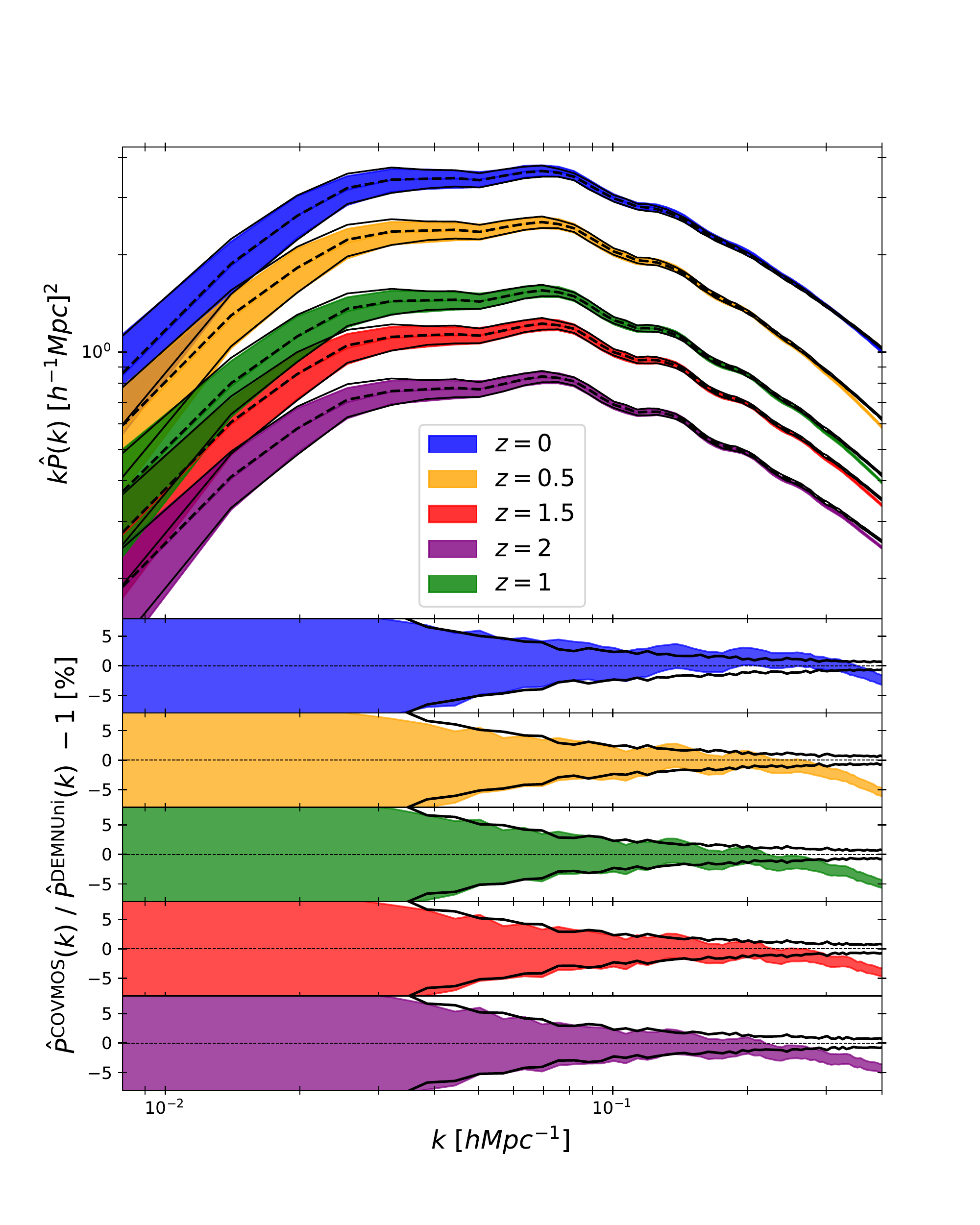}
\caption{\textit{Top panel}: Average of the estimated redshift-space power spectrum monopoles over $5,000$ realisations of \texttt{COVMOS} catalogues in the $16nu$ cosmology. The dispersion on each realisation is represented by the shaded area. The same quantities are represented in black for the $50$ \texttt{DEMNUni\_cov} realisations. \textit{Bottom panels}: The relative deviation between the averaged \texttt{COVMOS} and \texttt{DEMNUni\_cov} outputs with error bars.}
\label{fig:pkred}
\end{figure}

\begin{figure}
\hspace*{-0.6cm}\vspace*{-0.0cm}\includegraphics[scale=0.52]{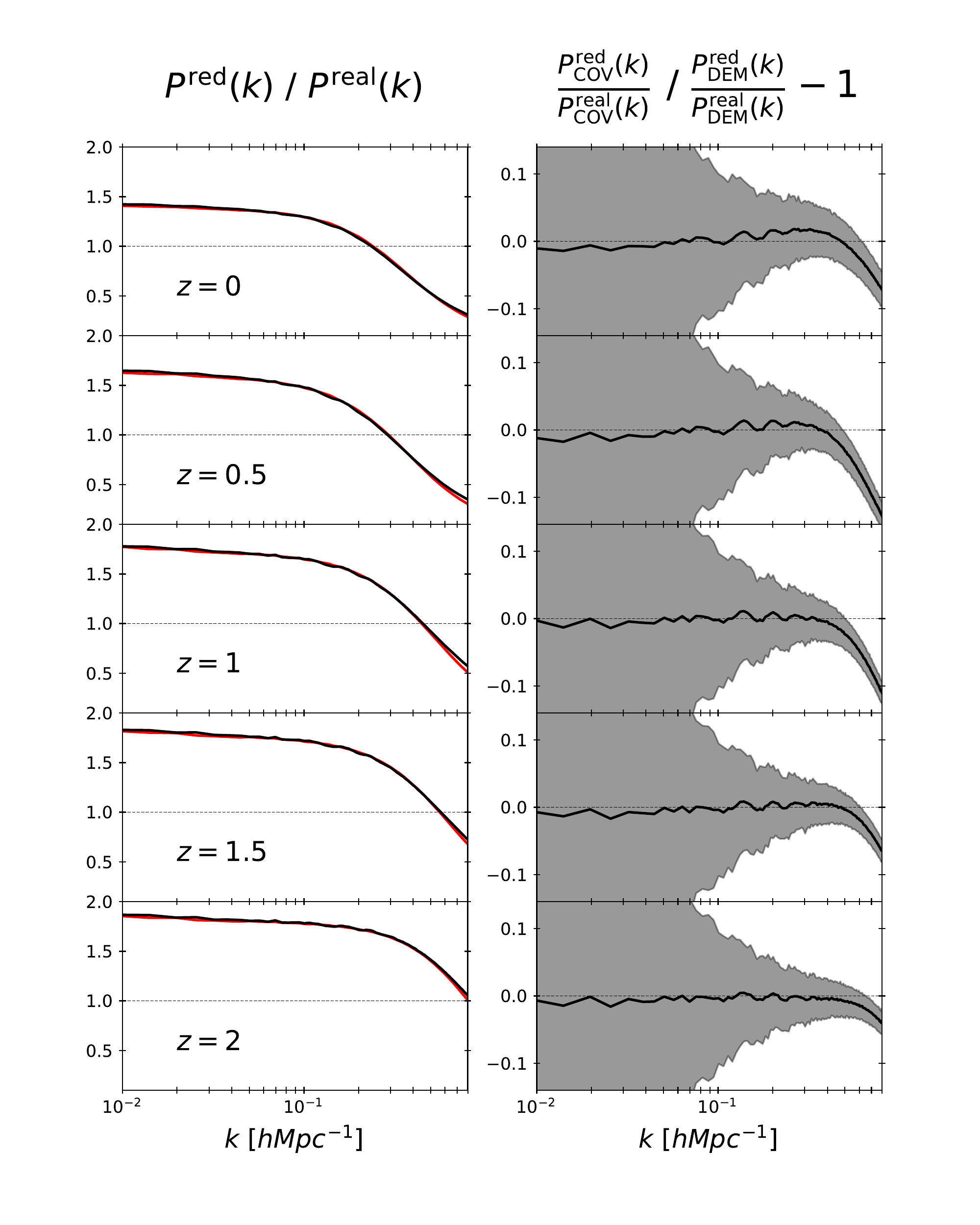}
\caption{\textit{Left}: ratio between averaged monopoles with and without RSD contribution in black for \texttt{DEMNUni\_cov} and in red for \texttt{COVMOS}, for the $5$ different redshifts. \textit{Right}: ratio between the previous quantities in black solid line with error bars in shaded grey, jointly accounting for estimated variance on \texttt{DEMNUni\_cov} and \texttt{COVMOS} in both real and redshift-space.}
\label{fig:redratio}
\end{figure}

In Figures \ref{fig:pkred} and \ref{fig:redratio} we compare the power spectra of \texttt{COVMOS} to those estimated in the reference \texttt{DEMNUni\_cov}, $N$-body simulations. First, when comparing Fig.~\ref{fig:pkred} to its real-space equivalent Fig.~\ref{fig:pkcomov}, no specific deterioration of the monopole can be detected. Instead, it appears that the clustering is better simulated in redshift-space, especially at small scales where the \texttt{COVMOS} power spectrum agrees with the $N$-body one up to a larger $k$ for most of the redshifts. 

In order to quantify the robustness of the proposed velocity model, one can see the redshift-space to real-space averaged monopole ratio in Fig.~\ref{fig:redratio}. In the left panels, we can see that the effects of RSD on the power spectrum are well reproduced in \texttt{COVMOS} when compared to \texttt{DEMNUni\_cov}. For both the enhancement of the apparent clustering in the linear regime ($\leq 0.1h/$Mpc) or its suppression in the mildly non-linear regime ($\geq 0.1h/$Mpc), the velocity model allows to reproduce with high fidelity the large scale limit (Kaiser boost) but also the transition to the small scales. 

Finally, looking at the ratio between the previous quantity in the \texttt{COVMOS} and \texttt{DEMNUni\_cov} cases allows to extract and discuss the validity domain of the proposed velocity modelling (right panel of Fig.~\ref{fig:redratio}). While the model seems to perfectly recover the RSD effects in the linear regime at the percent level, a gradual departure can be observed in the non-linear regime. We observe a lack of power for $k\geq 0.6\ h/$Mpc which can be due to the severe filtering of the \texttt{COVMOS} fields in this regime. Since one cannot expect the peculiar velocity assignment method to be adapted to highly suppressed clustering amplitude, the grid precision seems here not suited to validate the velocity model on highly non-linear scales ($k\geq 0.6\ h/$Mpc).

\begin{figure}
\hspace*{-1cm}\vspace*{-0.0cm}\includegraphics[scale=0.48]{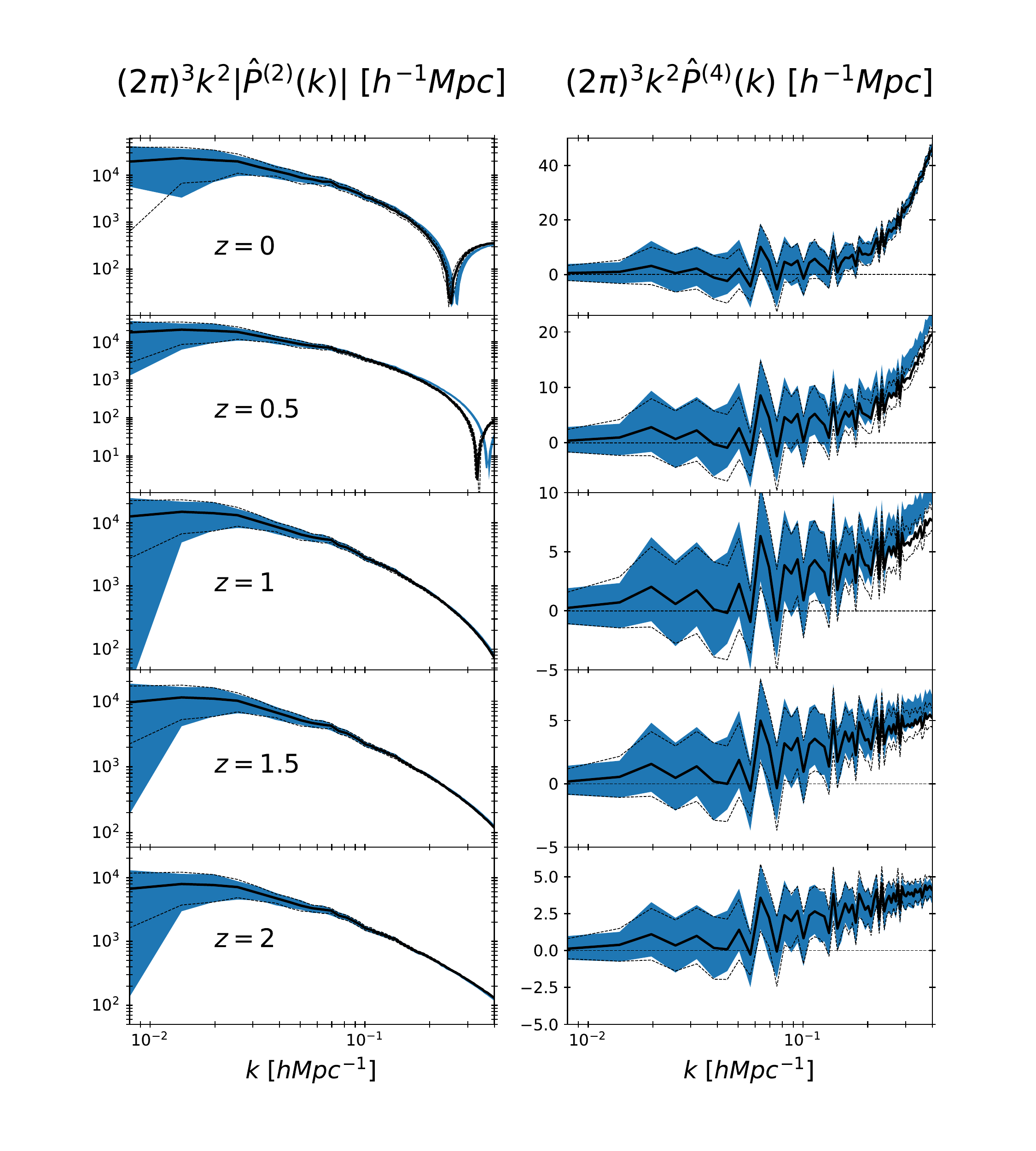}
\caption{Absolute values of the averaged quadrupoles (left panels) and averaged hexadecapoles (right panels) for the \texttt{DEMNUni\_cov} in solid black lines (error bars in dashed black lines) and for \texttt{COVMOS} with error bars in blue for the five redshifts.}
\label{fig:quadhexa}
\end{figure}

For completeness, in Fig.~\ref{fig:quadhexa} we compare the higher order multipoles of the power spectrum between \texttt{COVMOS} and \texttt{DEMNUni\_cov}. 
As in the case of the monopole, in the linear regime, both the amplitude and the dispersion of the simulated multipoles are in good agreement. In addition, it appears that in the mildly non-linear regime, these multipoles are not impacted by the various filterings affecting the monopoles (see Fig.~\ref{fig:pkred}). Indeed, none of these multipoles show a clear lack of power for $k>0.2~h/$Mpc, so that this anisotropic part of the power spectrum turns out to be reliably simulated on a wider range than the redshift-space monopole.

\begin{figure*}
\centering
\hspace*{-0.0cm}\vspace*{-0.0cm}\includegraphics[scale=0.5]{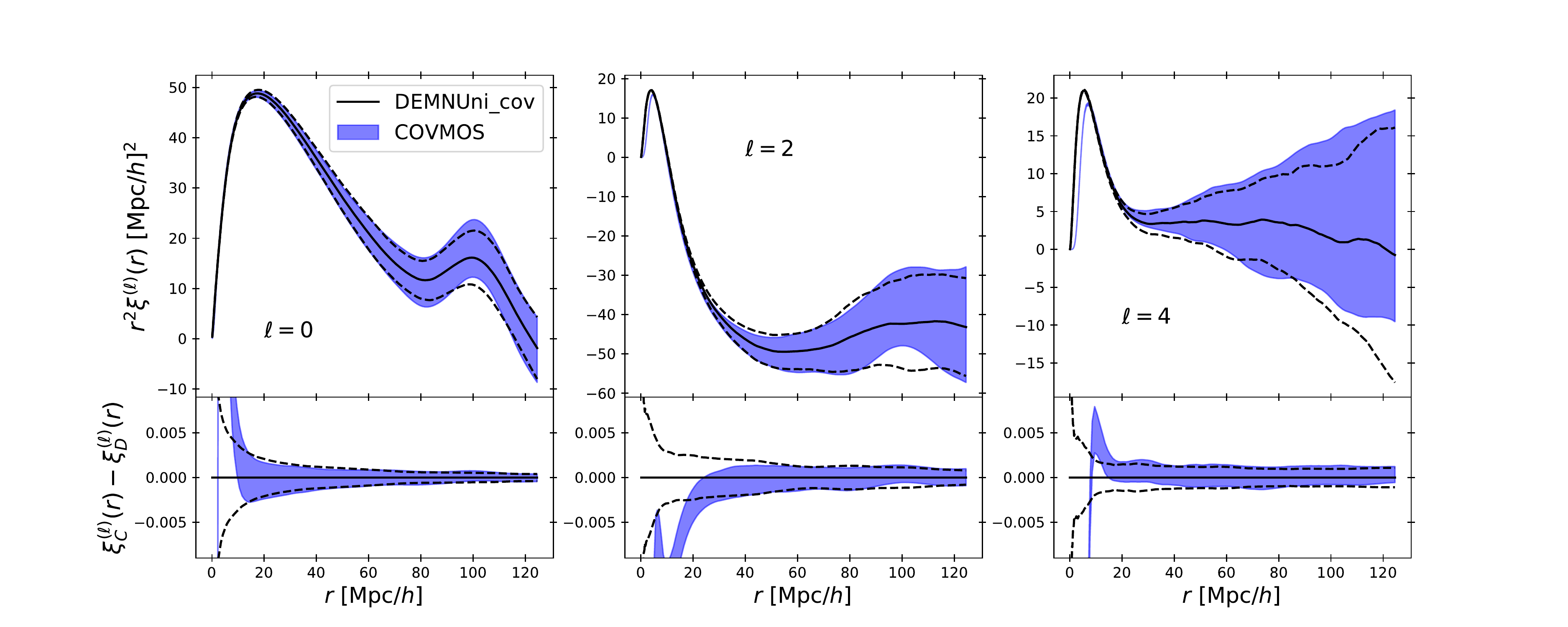}
\caption{\textit{Top}: Average  of the estimated $2$-correlation function multipoles over $100$ realisations of \texttt{COVMOS} catalogues in $16nu$ cosmology, at $z=0$ with dispersion on each realisation represented in blue. The solid black line represents the same (with dispersion over measurements in dashed black lines) for  \texttt{DEMNUni\_cov} simulations. \textit{Bottom}: The residual between the averaged \texttt{COVMOS} and \texttt{DEMNUni\_cov} multipoles with corresponding error bars.}
\label{fig:xired}
\end{figure*}

 As in real space, we also make a comparison in redshift-space at the level of the $2$-point correlation function. We ran the estimation of the configuration space correlation function multipoles (applying periodic boundary conditions) over a smaller sample of $100$ \texttt{COVMOS} realisations, in the $16nu$ cosmology only at $z=0$. In addition, we estimate the $2$-point correlation functions on the complete set of $50$ \texttt{DEMNUni\_cov} simulations.
Fig.~\ref{fig:xired} exhibits the averaged monopole, quadrupole and hexadecapole, compared between \texttt{COVMOS} and \texttt{DEMNUni\_cov} on the scale range $r\in[0,125]$ Mpc$/h$. One can see that the redshift space  multipoles produced by \texttt{COVMOS} are in agreement with what is expected from  \texttt{DEMNUni\_cov} at least down to 40 Mpc$/h$. Indeed, in that regime not only the amplitude is matching but also the dispersion of the measurements. One can notice, that despite the fact that we made this verification at redshift $z=0$, the monopole is well reproduced on even smaller scales ($\sim 20h^{-1}$Mpc) than the other multipoles. This is validating our scheme to assign velocities also in configuration space.

\subsection{Comparing the real and redshift-space covariance matrices}
In this section we compare the covariance matrices produced by the two methods, obtained from the standard estimator

\begin{align}
\hat C_{ij}& = \frac{1}{N-1}\sum_{s=1}^{N} [P^s(k_i)-\mu_i][P^s(k_j)-\mu_j]\ , \label{covmatrixestimator} \\
\mu_i &=\frac{1}{N}\sum_{s=1}^{N} P^s(k_i) \nonumber \ ,
\end{align}
where $P^s(k_i)$ is the power spectrum of the $s$-th simulation (over a total of $N$) evaluated at mode $k_i$.

This statistics, according to \cite{Scoccimarro:1999kp}, can be decomposed into two terms
\begin{equation}
C_{ij} = \frac{\hat P(k_i)^2}{M_{k_i}} \delta^K_{ij} +k_F^3 \bar T (k_i,k_j) \ ,
\label{Pkcovmatrixscoccimaro}
\end{equation}
respectively referred to as the Gaussian and non-Gaussian contributions. The former is related to the amplitude of the power spectrum, normalised by the number of independent averaged modes $M_{k_i}$ within shells centred around $k_i$, only affecting the diagonal of the matrix. The second affects the whole matrix and depends on the cross shell-averaged trispectrum 
\begin{align}
    \bar T(k_i,k_j) = \int_{k_i}\int_{k_j} T(\vec k_1,-\vec k_1,\vec k_2,-\vec k_2) \frac{\mathrm d^3 \vec k_1}{V_{k_i}}\frac{\mathrm d^3 \vec k_2}{V_{k_j}}\ ,
    \label{binavgTri}
\end{align}
where the trispectrum can be constructed from the density field as
\begin{align}
    \left< \delta_{\vec {k_1}}\delta_{\vec {k_2}}\delta_{\vec {k_3}} \delta_{\vec {k_4}}\right>_c = \delta^D(\vec {k_1}+\vec {k_2}+\vec {k_3}+\vec {k_4})\ T(\vec {k_1},\vec {k_2},\vec {k_3})\ .
\end{align}
Expected to be zero in the case of a Gaussian realisation, this second contribution directly weighs the non-Gaussian nature of the simulated cosmic fields.

While the first term in the case of the \texttt{COVMOS} method is expected to be suppressed when the various filterings start to act ($k>0.2h/$Mpc for a grid precision of $N_s=1024$), the second term is not so obvious to guess. The main approximation of \texttt{COVMOS} is to suppose that $\bar T(k_i,k_j)$ can be to some extent properly reconstructed if both the power spectrum and the PDF are accurate. However such a connection remains difficult to foresee. In any case, this term is expected to dominate the non-linear regime and to raise in power as the targeted PDF follows a more clustered, skewed shape.

To the estimator of the covariance Eq.~\ref{covmatrixestimator} will be associated in the following an analytical error-bar, predicted when assuming a Gaussian multivariate distribution of the data \citep{10.2307/2331939}:
\begin{equation}
    V\left[ \hat C_{ij}\right] = \frac{C_{ij}^2+C_{ii}C_{jj}}{N-1}\ .
    \label{gauss_error_on_cov}
\end{equation}

\begin{figure*}
\centering
\includegraphics[width=1.\linewidth]{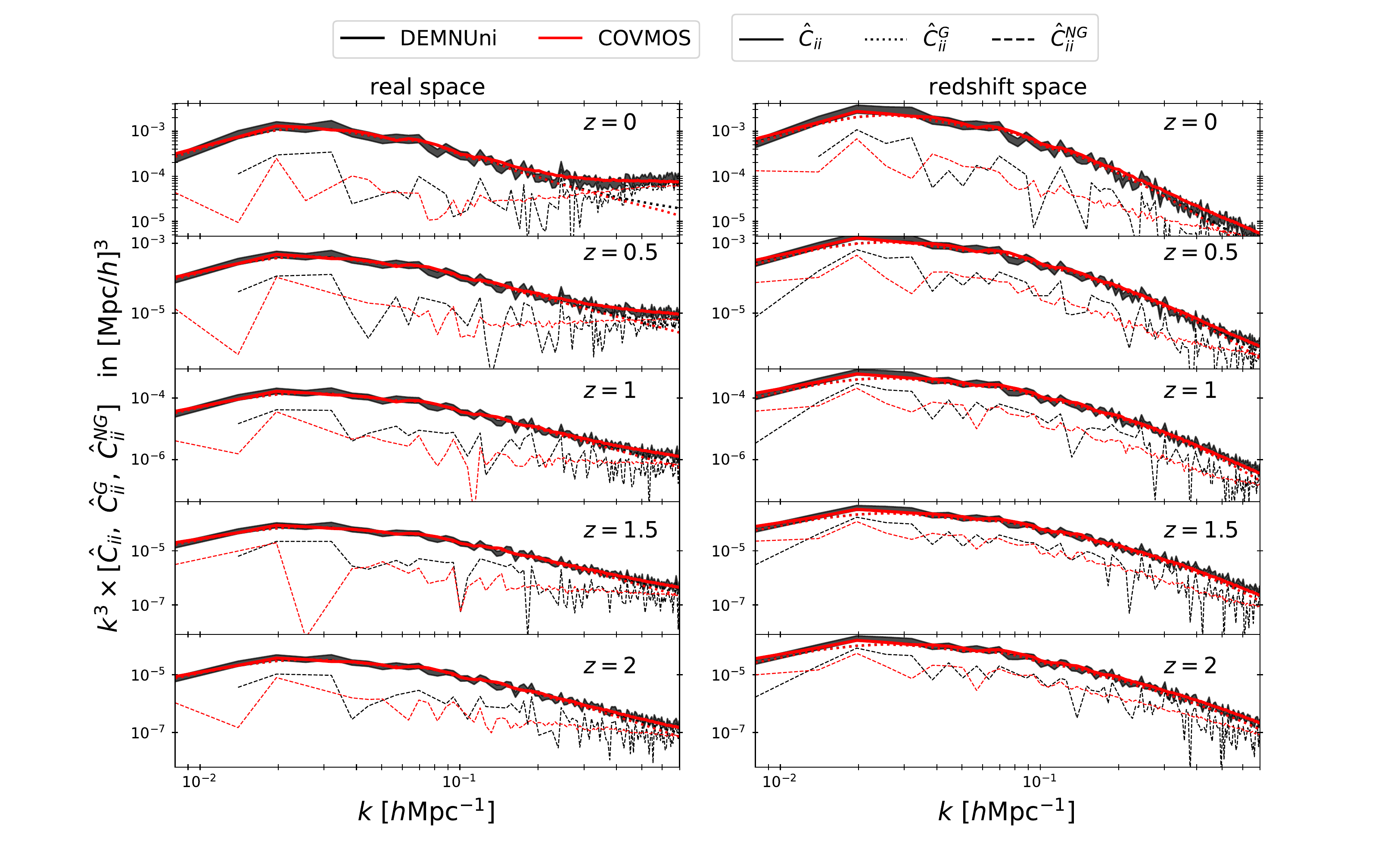}
\caption{Estimated contributions of the diagonal of the covariance matrices over $5,000$ realisations of \texttt{COVMOS} catalogues (in red) in real (left) and redshift-space (right) as compared to the ones estimated from $50$ \texttt{DEMNUni\_cov} (in black) for the five redshifts and in $16nu$ cosmology. The estimated diagonals $\hat C_{ii}$ with Gaussian error bars (see Eq.~\ref{gauss_error_on_cov}) are represented in solid lines. In dotted lines are displayed the estimated Gaussian contribution $\hat C_{ii}^\mathrm{G}$ of the covariance matrix estimated from the averaged power spectra (see Eq.~\ref{Pkcovmatrixscoccimaro}). The estimated non-Gaussian contributions $\hat C^\mathrm{NG}_{ii}$ are drawn in dashed lines and are coming from the residual between the full diagonal and the Gaussian term alone.}
\label{fig:diagred}
\end{figure*}

Note that when using the standard estimator of the covariance Eq.~\ref{covmatrixestimator}, the \texttt{COVMOS} pipeline leads to a systematic over-estimate of the covariance. Such a bias can be identified and has been subtracted in the following; the details of this procedure can be found in appendix \ref{debiaisingcovmatrix}.

In Fig.~\ref{fig:diagred} are shown the contributions to the variance, both in the \texttt{DEMNUni\_cov} and the \texttt{COVMOS} cases, for all redshifts and in the $16nu$ cosmology. 
The figure shows that for the five redshifts, the full variances $\hat C_{ii}$ are matching within Gaussian error bars. This observation can also be extended to modes well beyond those for which the power spectrum is correctly reproduced. This effect can be understood when decomposing the variance in Gaussian and non-Gaussian contributions (Eq.~\ref{Pkcovmatrixscoccimaro}): the former is computed from the average power spectrum over the realisations whereas the second is obtained subtracting the Gaussian contribution from the total variance. In doing so, one can see that in the linear regime, the Gaussian term that dominates about one order of magnitude the second term unsurprisingly follows the same trend as discussed for the power spectrum. It is accurately reproduced but starts to disagree with the $N$-Body reference around $k=0.2h/$Mpc. On the other hand, as the damping happens at scales where the trispectrum term starts to dominate, it allows to recover this lack of power and produces an overall variance with correct amplitude. Nevertheless, we can observe at $z=0$ a systematic overestimate of this term, from $k\sim 0.25h/$Mpc both in real and redshift-space. This could be due to the insufficiency of our approximations at very low redshift.

\begin{figure*}
\vspace{-0.4cm}
\subfloat{\includegraphics[width=0.94\linewidth]{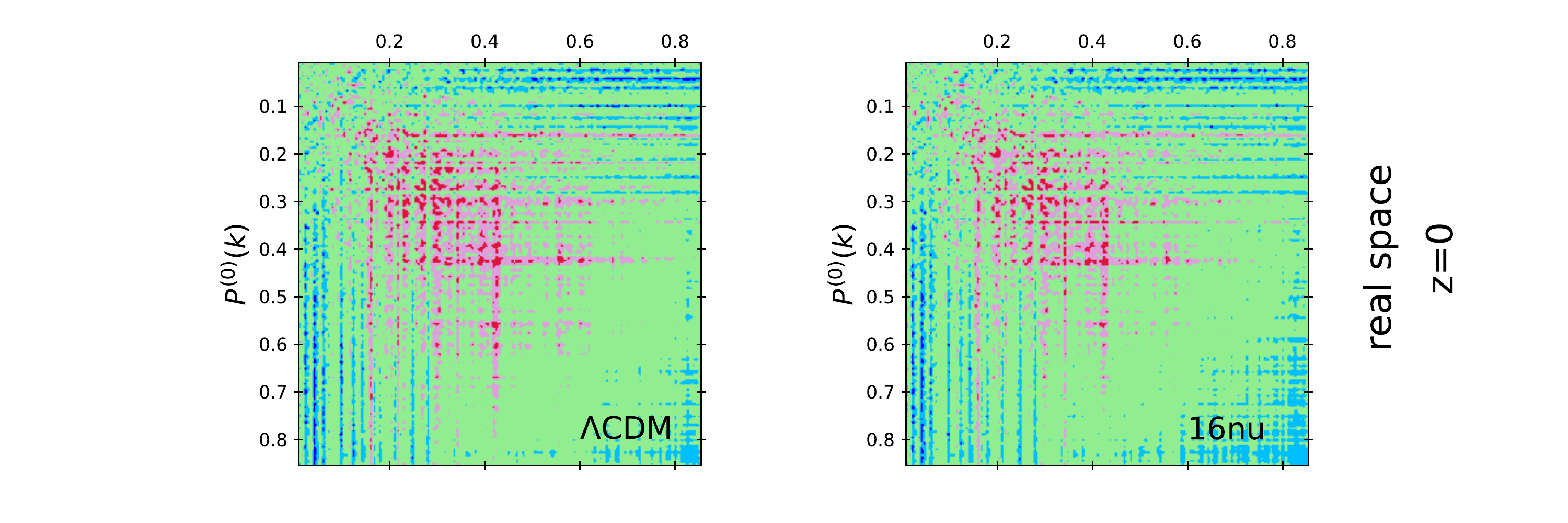}}
\vspace{-0.75cm}
\subfloat{\includegraphics[width=0.94\linewidth]{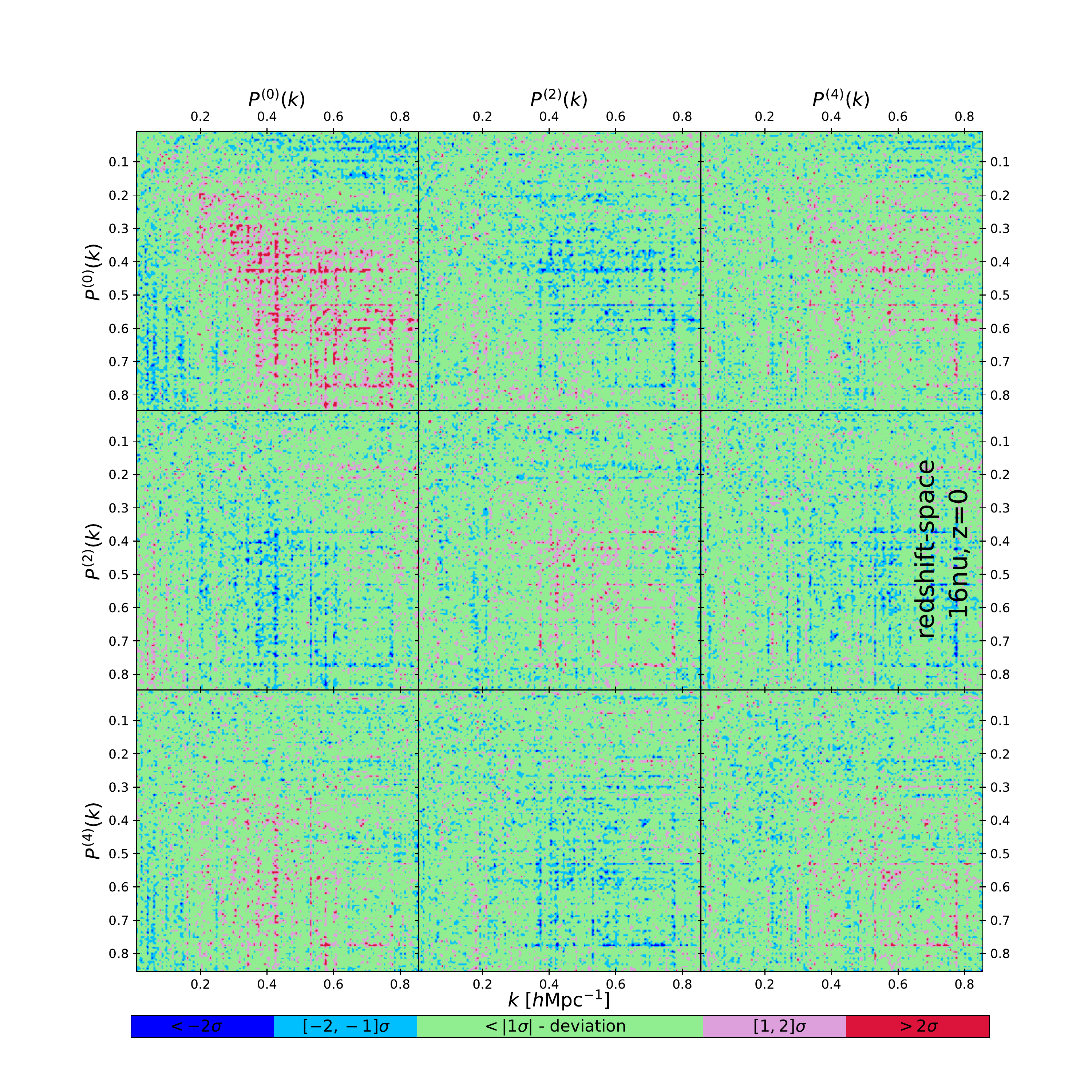}}
\caption{Difference in $\sigma$-deviation between the cross-covariance matrices of the power spectrum in Fourier space obtained from $5000$ \texttt{COVMOS} and $50$ \texttt{DEMNUni\_cov} realisations at $z=0$. This difference is normalised by the predicted Gaussian error Eq.~\ref{gauss_error_on_cov}, see Eq.~\ref{residual_normalised}. The upper panels are given for the monopoles in real space for the two cosmologies, while the bottom panels stand for the cross-covariance terms in redshift-space, for the $16nu$ cosmology only.}
\label{fig:cov_z0}
\end{figure*}

\begin{figure*}
\vspace{-0.4cm}
\subfloat{\includegraphics[width=0.94\linewidth]{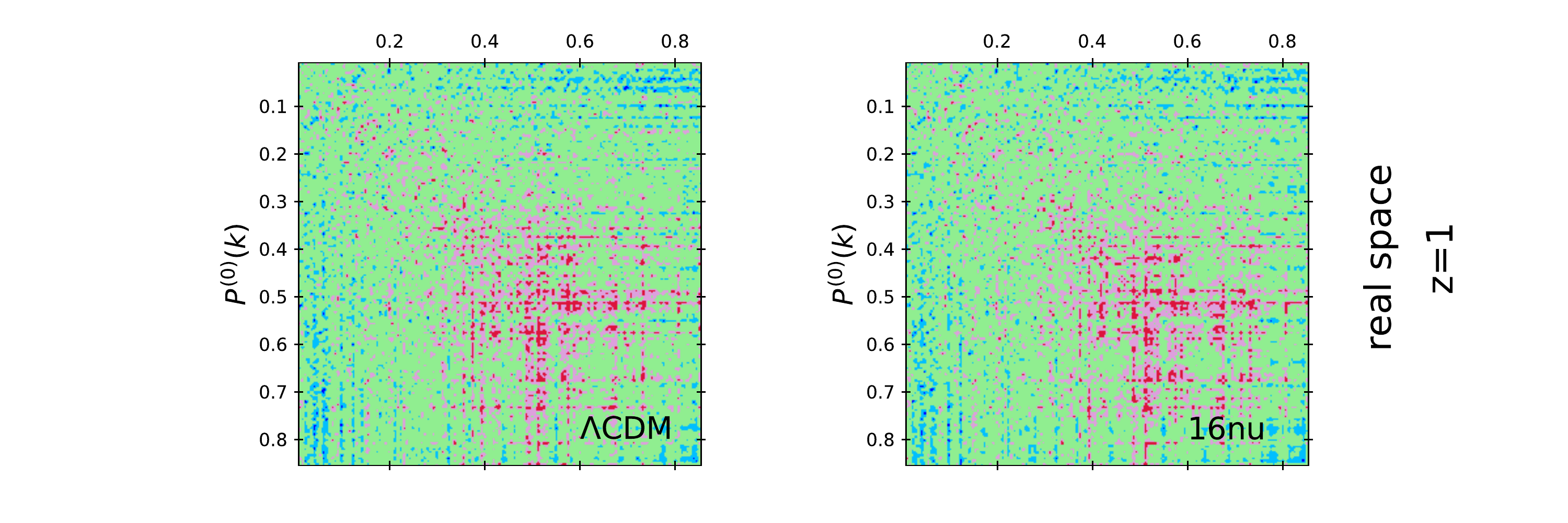}}
\vspace{-0.75cm}
\subfloat{\includegraphics[width=0.94\linewidth]{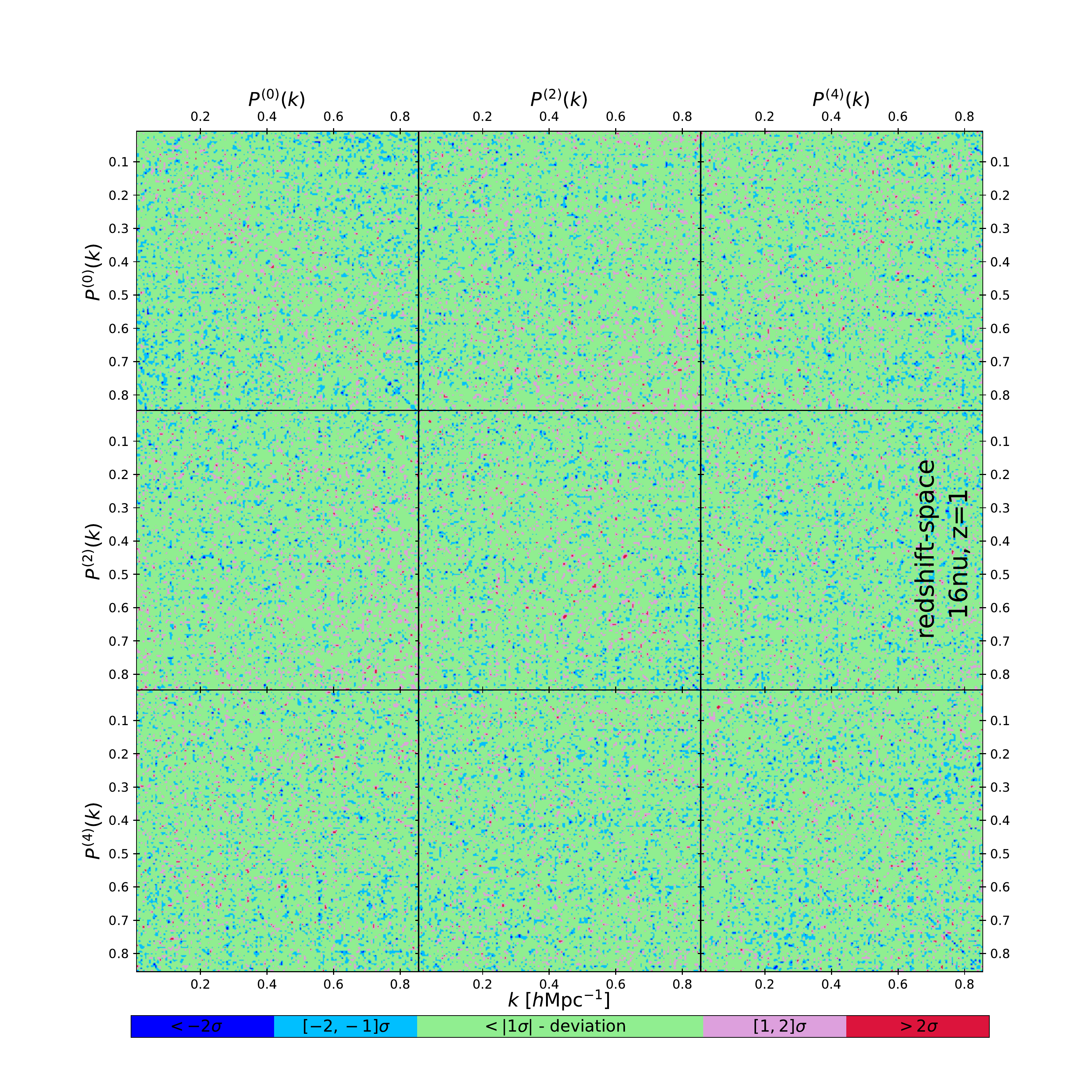}}
\caption{Difference in $\sigma$-deviation between the cross-covariance matrices of the power spectrum in Fourier space obtained from $5000$ \texttt{COVMOS} and $50$ \texttt{DEMNUni\_cov} realisations at $z=1$. This difference is normalised by the predicted Gaussian error Eq.~\ref{gauss_error_on_cov}, see Eq.~\ref{residual_normalised}. The upper panels are given for the monopoles in real space for the two cosmologies, while the bottom panels stand for the cross-covariance terms in redshift-space, for the $16nu$ cosmology only.}
\label{fig:cov_z1}
\end{figure*}

\begin{figure*}
\vspace{-0.4cm}
\subfloat{\includegraphics[width=0.94\linewidth]{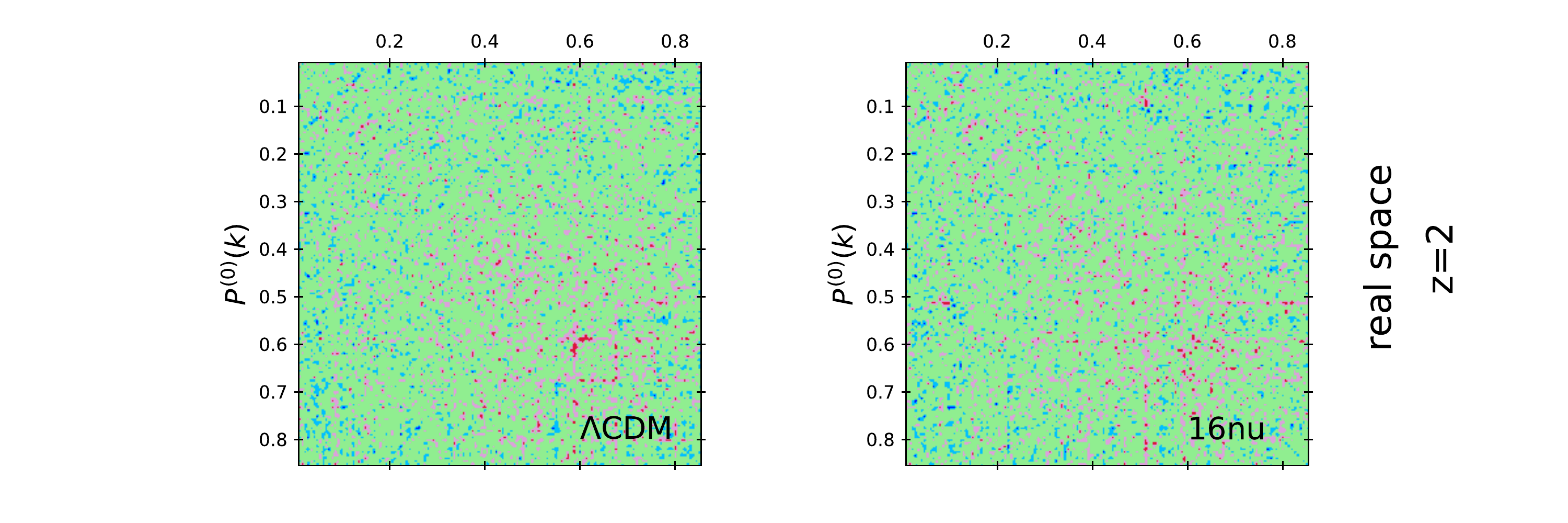}}
\vspace{-0.75cm}
\subfloat{\includegraphics[width=0.94\linewidth]{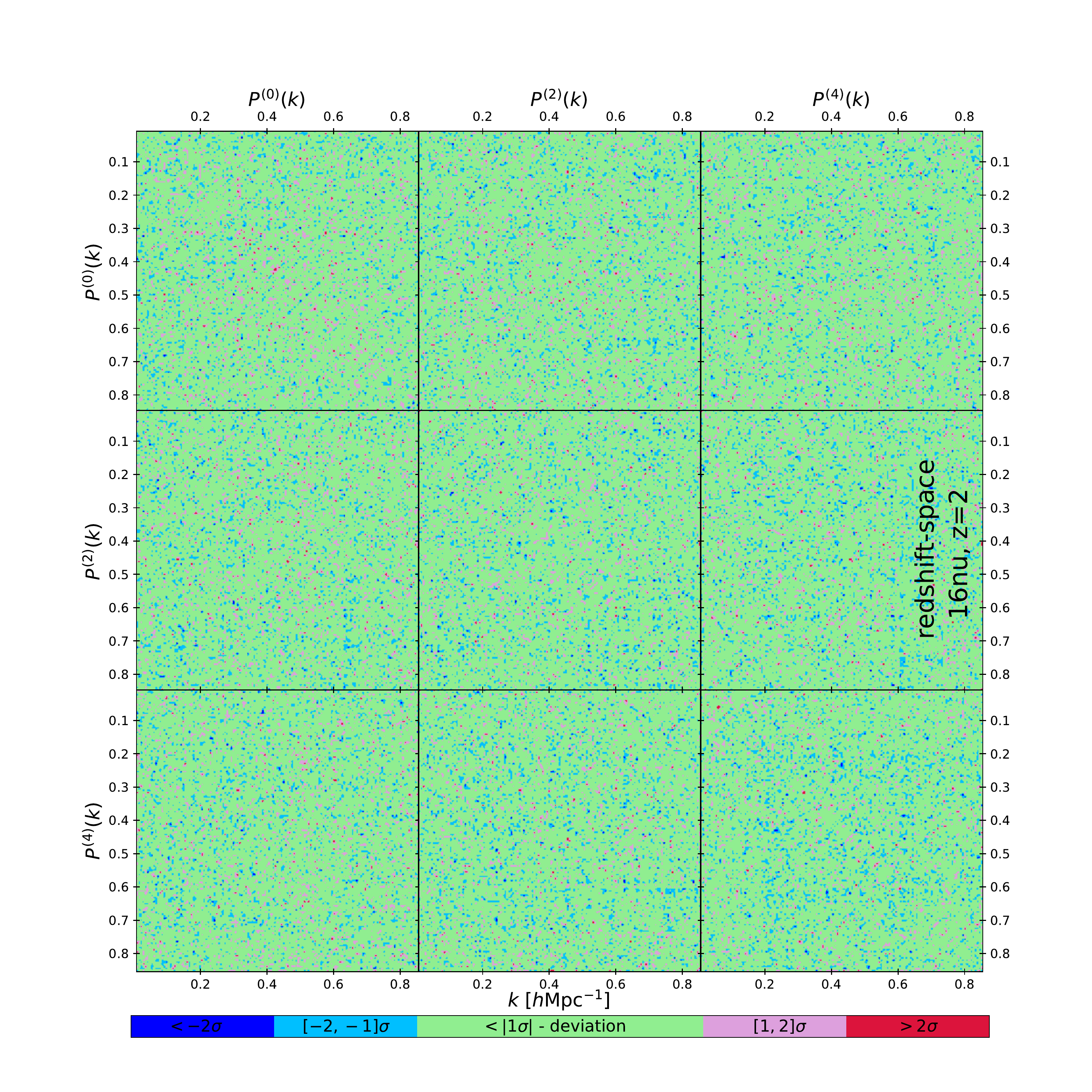}}
\caption{Difference in $\sigma$-deviation between the cross-covariance matrices of the power spectrum in Fourier space obtained from $5000$ \texttt{COVMOS} and $50$ \texttt{DEMNUni\_cov} realisations at $z=2$. This difference is normalised by the predicted Gaussian error Eq.~\ref{gauss_error_on_cov}, see Eq.~\ref{residual_normalised}. The upper panels are given for the monopoles in real space for the two cosmologies, while the bottom panels stand for the cross-covariance terms in redshift-space, for the $16nu$ cosmology only.}
\label{fig:cov_z2}
\end{figure*}

\begin{figure*}
\vspace{-0.4cm}
\subfloat{\includegraphics[width=0.94\linewidth]{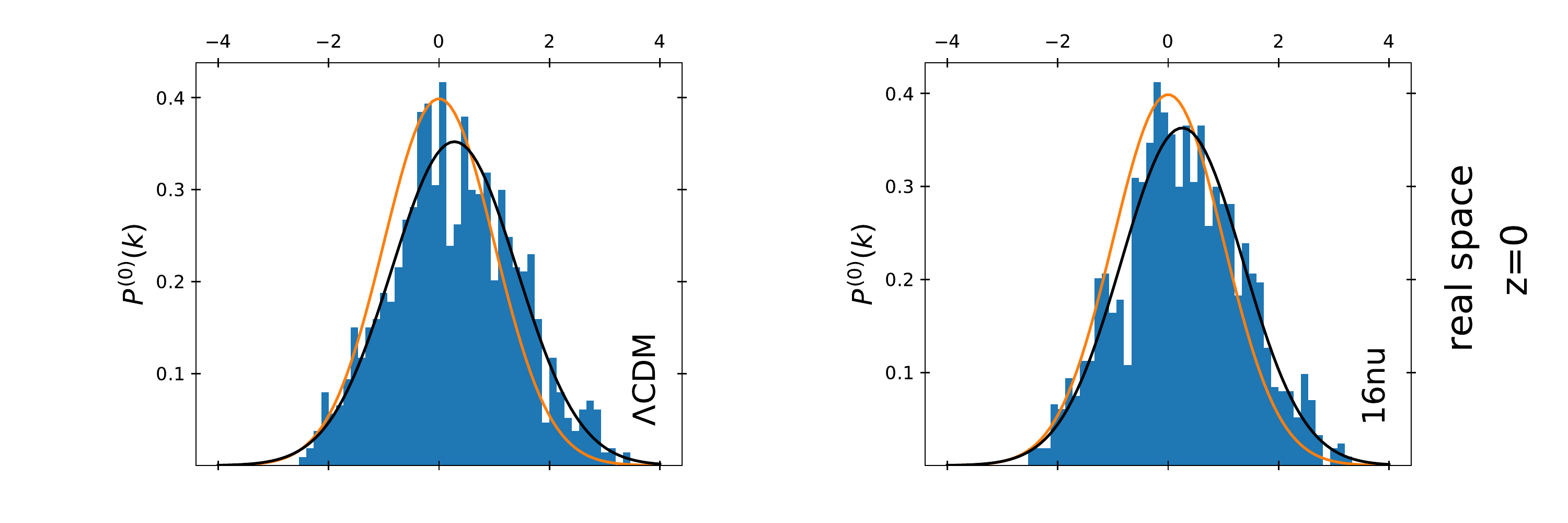}}
\vspace{-0.75cm}
\subfloat{\includegraphics[width=0.94\linewidth]{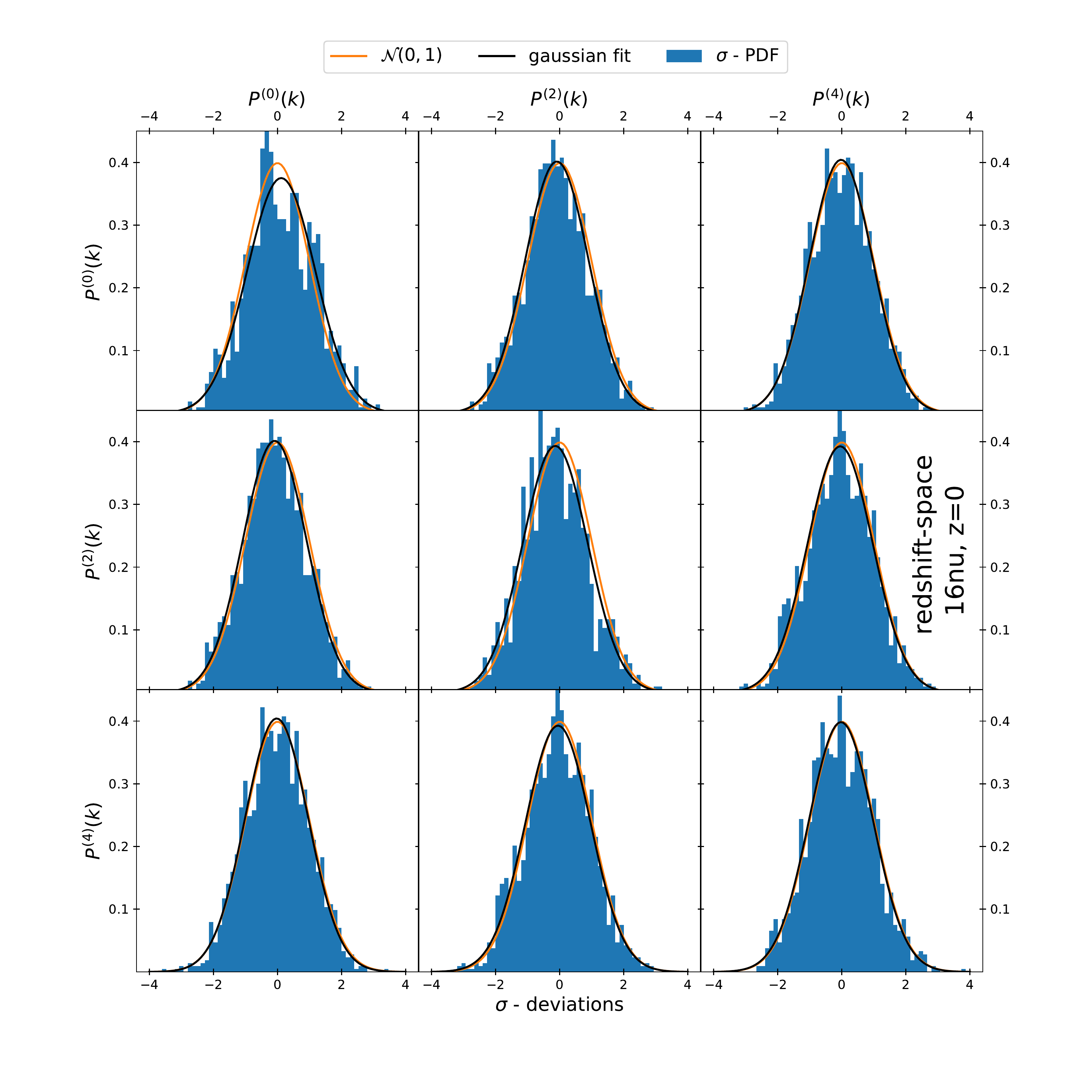}}
\caption{Normalised distributions of the residual of the covariance matrices normalised by Gaussian error (the $\sigma$-deviation distribution, see Fig.~\ref{fig:cov_z0}) for the multipoles of the power spectrum at $z=0$ in blue. Its Gaussian fit (on the two first moments) is represented in black and is compared to the standardised normal distribution in orange. The upper panels presents the distributions in real space for the two cosmologies, while the bottom panels give the results in redshift-space, for the $16nu$ cosmology only. In these panels, the maximal Fourier mode is set at $k_{max} = 0.25h/$Mpc.}
\label{fig:pool}
\end{figure*}

We compare all terms of the covariance matrices obtained for the $N$-body and the \texttt{COVMOS} method for the three redshifts $z=[0,1,2]$ in Figures \ref{fig:cov_z0}, \ref{fig:cov_z1} and \ref{fig:cov_z2}, respectively. This comparison is conducted in real-space for the monopole, and in redshift-space for all multipoles, where we also show the cross-covariances between multipoles. We represent the residual of the \texttt{COVMOS} and \texttt{DEMNUni\_cov} matrices in standard deviation, i.e.

\begin{equation}
    R_{ij} = \frac{\hat C_{ij}^{COVMOS} - \hat C_{ij}^{DEMNUni}}{\sqrt{{\sigma^2}_{ij}^{COVMOS}+{\sigma^2}_{ij}^{DEMNUni}}}\ , 
    \label{residual_normalised}
\end{equation}
 where the $\sigma^2_{ij}$ terms are computed using Eq.~\ref{gauss_error_on_cov}.
Looking at these three figures, one can first remark a large number of over or under-estimate of the covariance terms without apparent global trend. They constitute the only observable feature that one can see in redshift-space at $z=1$ and $z=2$ and in real space at $z=2$. This is explained by the stochastic noise, unrelated to the validity of the method, expected to follow a centered and reduced normal distribution. Thus drawing the distribution of $R_{ij}$ against its prediction helps to discuss the validity of the produced covariances. To lighten the amount of figures, this is shown only at $z=0$ in Fig.~\ref{fig:pool}, taking a maximal mode $k_{max} = 0.25h/$Mpc. In this regime in redshift-space, one can see that the measured distribution is in good agreement with the prediction, demonstrating that no bias in the \texttt{COVMOS} matrices can be detected. This is not as strong in real space, where one can detect a shift of the distribution toward larger $\sigma$ values. This depicts an overall over-estimate of the covariance elements, that can be easily spotted in the top panels of Fig.~\ref{fig:cov_z0}.

Second, one can see in real space a migration of an over-correlation pattern toward higher scales for decreasing redshift. These over-correlations are maximised around the diagonal at $z=0$ from $k\sim 0.2h/$Mpc to $k \sim 0.5h/$Mpc and no change can be noticed when switching from one cosmology to another.
Once again, this can be attributed to the weak validity of the \texttt{COVMOS} approximations at low redshift $z\sim 0$.

On the other hand, it turns out that the redshift-space covariance terms are systematically better simulated than their equivalents in real space. This effect is more apparent at $z=0$ where the systematic over-estimate of the correlations in real space are significantly reduced in redshift-space. This improvement is in agreement with Fig.~\ref{fig:diagred}, showing a slight increase of the range of reliability of the trispectrum in redshift than in real space. According to Eq.~\ref{Pkcovmatrixscoccimaro}, the off-diagonal terms of the covariance matrix are only related to the trispectrum terms, explaining why the whole matrices in redshift-space are more accurate than in real-space.

Concerning the other multipoles and their cross-covariance terms in redshift-space, no particular over or under-correlation pattern throughout the simulated redshifts can be observed beyond the noise, as depicted in Fig.~\ref{fig:pool}. At these probed redshifts, both error bars and correlations of the multipoles are in good agreement with the $N$-body outputs.

\begin{figure*}
\centering
\hspace*{-0.0cm}\vspace*{-0.0cm}\includegraphics[scale=0.55]{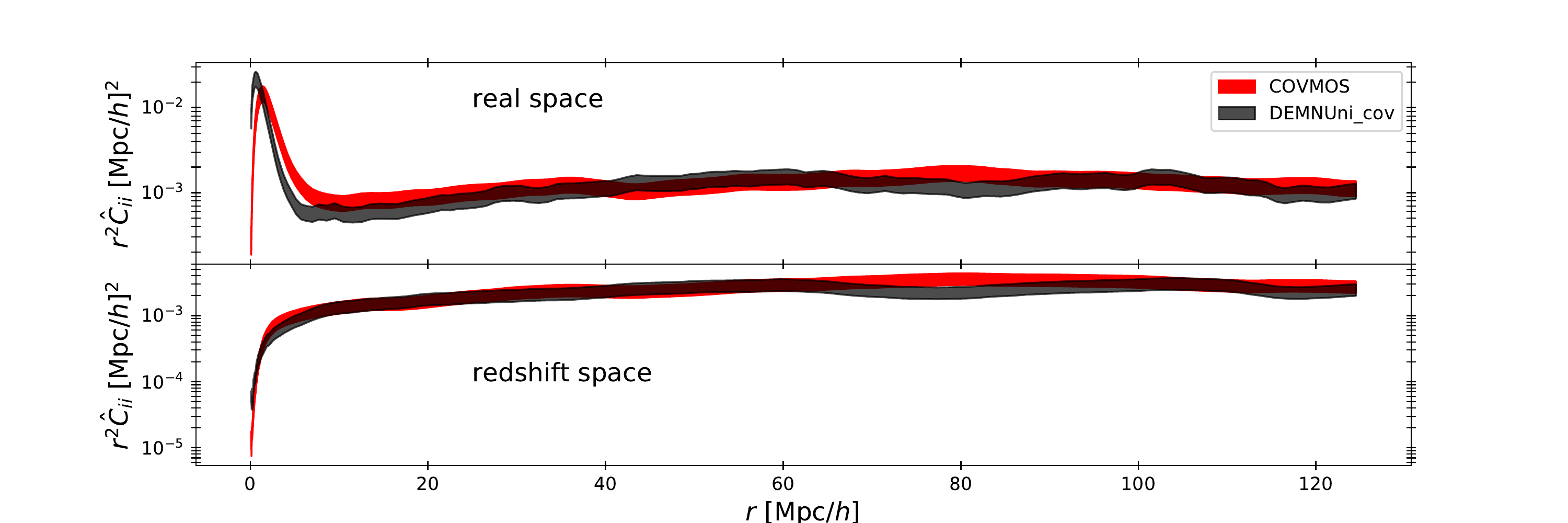}
\caption{Estimated diagonal of the $2$-point correlation function covariance matrix in real (top panel) and redshift-space (bottom panel) estimated from $50$ \texttt{DEMNUni\_cov} realisations in black and $100$ \texttt{COVMOS} realisations in red. Assuming a Gaussian distribution of the data, errors bars are computed from Eq.~\ref{gauss_error_on_cov}.}
\label{fig:CiiXi}
\end{figure*}

\begin{figure*}
\vspace{-0.4cm}
\subfloat{\includegraphics[width=0.94\linewidth]{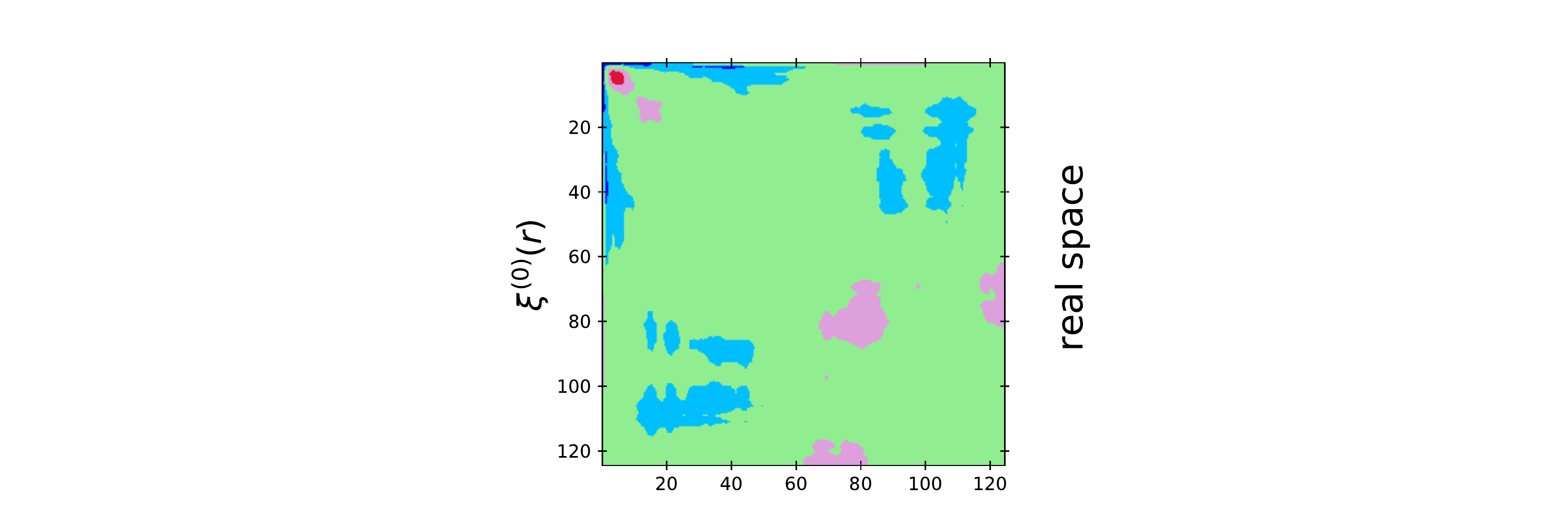}}
\vspace{-0.75cm}
\subfloat{\includegraphics[width=0.94\linewidth]{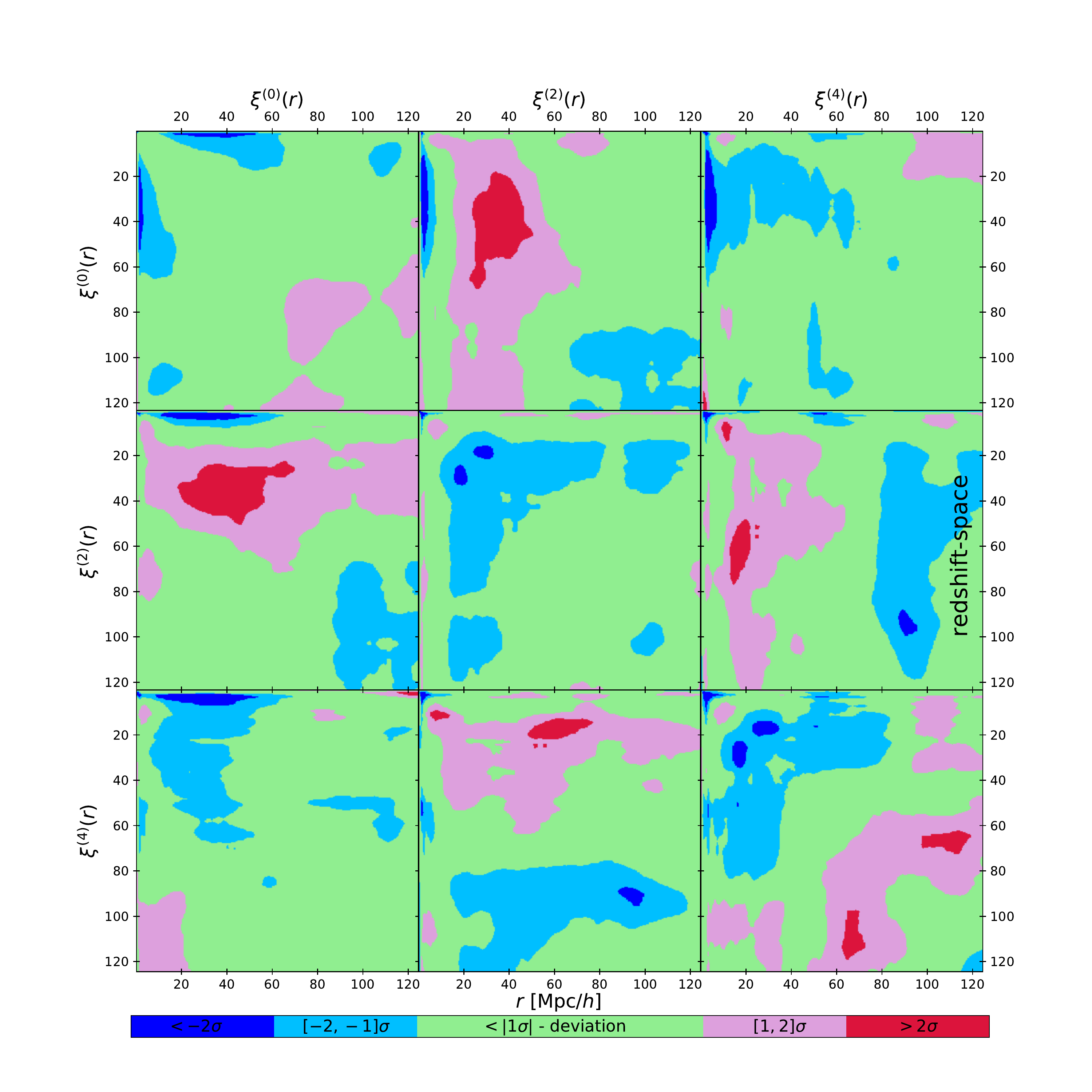}}
\caption{Difference in $\sigma$-deviation between the covariance matrices of the multipoles of the $2$-point correlation function estimated from $100$ \texttt{COVMOS} realisations and $50$ \texttt{DEMNUni\_cov} realisations. This difference is normalised by the predicted Gaussian errors estimated from Eq.~\ref{gauss_error_on_cov}. The upper panel stands from the comparison of the monopole in real space, while the bottom panels stands for the covariance terms of the multipoles in redshift-space.}
\label{fig:Cij_Xi}
\end{figure*}

Finally, Fig.~\ref{fig:CiiXi} and Fig.~\ref{fig:Cij_Xi} are respectively showing the results for the variance and the covariance of the 2-point correlation function multipoles, both in real and redshift-space.

Focusing on real space, the variance of the $2$-point correlation function monopole seems to be reproduced in a reliable way down to $\sim 6$ Mpc$/h$, before being successively over- ($2<r<6$ Mpc$/h$) and under- ($r<2$ Mpc$/h$) estimated, which is in agreement with the observation made directly on the $2$-point correlation function monopole in Fig.~\ref{fig:xired}. These observations can also be made for the whole monopole covariance in the upper panel of Fig.~\ref{fig:Cij_Xi}: a majority of matrix elements fall in the $<|1\sigma|$ deviation ($>68\%$) in the regime $>\sim 6$ Mpc$/h$ before being over and under-estimated above this limit.

In redshift space, as for the power spectrum analysis, a significant improvement in the simulation of the errors and correlations are observed in the small scales. The variance of the monopole is now reliable down to $\sim 3$ Mpc$/h$, while the under-estimation of the correlation between scales which was observed in real space is damped in this case.
Despite the fact that we have a less significant number of realisations in the case of the $2$-point correlation function one can see that in general the covariance and cross-covariance of the $2$-point correlation multipols is well reproduced. More realisations would be needed to assess the significance of the $2$-sigma deviations around $30$ Mpc$/h$ in the cross-covariance between the monopole and the quadrupole and in the covariance of the quadrupole. Indeed, we started with $50$ realisations only and generating the $50$ more greatly improved the comparison.

\section{Conclusion and discussion}
\label{conclusion}

Although apparently straightforward, relying on covariance matrices for $2$-point statistics estimators evaluated from mock catalogues generally leads to two main concerns. The first is the sampling noise that propagates down to the parameter constraints and turns out to be difficult to correct without making any assumption on the posterior distribution of parameters \citep{Sylvainpaper, dodelson_13, percival_13, taylor_13, taylor_14}. The second, and less investigated, is the computational cost of evaluating a covariance matrix for various set of cosmological parameters in order to study potential biases due to the fiducial choice of the cosmology when analysing data.

In this context, we proposed a method implemented in the public code named \texttt{COVMOS}, allowing the fast production of a large number
of catalogues. Each individual catalogue is a realisation of a non-Gaussian density field, characterised by a probability density function and a power spectrum.  In~\citet{Baratta:2019bta}, we showed that it is possible to reproduce with a high level of accuracy a given PDF and a power spectrum, but it was limited to real space and to a Log-Normal PDF. In the present paper, we present two major improvements :
\begin{itemize}
    \item the probability distribution function of the density field can be arbitrarily set as an input, this way one can mimic the PDF measured from a N-body simulation or provide a theoretical model,
    \item we designed a peculiar velocity assignment procedure which accurately models the effect of redshift-space distortion on $2$-point statistics at least up to $k = 0.5 h$ Mpc$^{-1}$.
\end{itemize} 
As a result, one needs to specify a set of statistical quantities and relations as inputs of the code. These are used as targets for the pipeline. These include (i) the power spectra of the density and  (ii) velocity fields, (iii) the one-point probability distribution function of the density field, and (iv) a relation between the velocity dispersion and the density.

In addition, we validated the whole \texttt{COVMOS} pipeline in real and redshift-space, for $2$-point statistics (the power spectrum and the correlation function) and their associated covariance matrices, against state-of-the art $N$-body simulations (DEMNUni). We investigated performances for two cosmological models, a standard $\Lambda$CDM model and one including massive neutrinos. In both cosmological scenarios, we showed that for a spatial resolution of the order of $1h^{-1}$Mpc, in a comoving volume of $1h^{-3}$Gpc, the matter power spectrum multipoles in redshift space were reliably reproduced up to $k = 0.25\ h/$Mpc (down to $r=10h^{-1}$Mpc for the $2$-point correlation function). 

We extended the validation of the method to $4$-point statistics in Fourier space. More precisely, we verified that the shell-averaged trispectrum, $\bar T$, which enters the power spectrum covariance, is in good agreement with the one obtained from $50$ $N$-body simulations. This resulted in an accurate reproduction of the power spectrum covariance matrices. Although, slightly overestimated in real space for $k \sim 0.15\ h/$Mpc, it turns out that in redshift space we can reach $k \sim 0.2\ h/$Mpc for the monopole power spectrum.
 
Having validated the method against $N$-body simulation, the major advantage of the code is its numerical efficiency in terms of computational cost. As an example, for this work we generated $5\times 5000$ catalogues of $10^8$ objects in $20$ days on a  HPC cluster\footnote{The Dark Energy Cluster (DEC) hosted in CPPM, Marseille.}, using $940$ processors of $2.4$ GHz. Thus, we are able to generate an extremely large number (25k) of realisations in less than 20k cpu-hr, in order to produce noise-free covariance matrices. Indeed, it is faster to generate a realisation than estimating the power spectrum of that realisation, and the difference is even more important when considering the estimation of the $2$-point correlation function. 

The method may need to be further investigated to reach a higher level of precision and accuracy before a potential application in the final data analysis of large galaxy surveys. However, the associated public code can already be useful for a large variety of applications. Indeed, its ability to simulate a wide variety of cosmological models and to achieve almost no sampling noise in the produced covariance matrices makes this method particularly relevant for testing Likelihood pipelines \citep[see][for a first application]{Sylvainpaper} aiming at constraining cosmological parameters. 

Finally, we recall that the \texttt{COVMOS} code is public \footnote{\href{https://github.com/PhilippeBaratta/COVMOS}{github.com/PhilippeBaratta/COVMOS}}, fully written in Python, ready to be used and is adapted to receive the needed inputs either from theory or simulations.

A natural extension of this method, which would get it closer to observations, would be the reconstruction of the observed light-cone, in order to account, e.g., for survey window function effects. In addition, this would allow the production of angular maps, suited for the study of photometric statistics in the Spherical Harmonics domain. Being a method specialised in $2$-points statistics, correlated maps
of observables such as galaxy clustering and galaxy shear, could be produced, as well as their associated cross-covariances. Future extensions could also include radio observations like CMB temperature maps, CMB-lensing or even $21$-cm maps. Given the increase in their performance that the exploitation of cross-correlations between different observables are and will provide to ongoing and future cosmological surveys, especially for parameter degeneracy breaking through their combinations, the \texttt{COVMOS} method and further developments could definitely be useful for modern cosmology.

\begin{acknowledgements}
PB and SGB were supported by CNES, focused on Euclid mission. The project leading to this publication has received funding from Excellence Initiative of Aix-Marseille University -A*MIDEX, a French "Investissements d'Avenir" programme (AMX-19-IET-008 -IPhU). The DEMNUni simulations were carried out in the framework of ``The Dark Energy and Massive-Neutrino Universe" project, using the Tier-0 IBM BG/Q Fermi machine and the Tier-0 Intel OmniPath Cluster Marconi-A1 of the Centro Interuniversitario del Nord-Est per il Calcolo Elettronico (CINECA). We acknowledge a generous CPU and storage allocation by the Italian Super-Computing Resource Allocation (ISCRA) as well as from the coordination of the ``Accordo Quadro MoU per lo svolgimento di attività congiunta di ricerca Nuove frontiere in Astrofisica: HPC e Data Exploration di nuova generazione'', together with storage from INFN-CNAF and INAF-IA2.
\end{acknowledgements}
\bibliographystyle{aa}
\bibliography{bibliography.bib}

\appendix
\section{Debiasing the Monte Carlo covariance matrices}
\label{debiaisingcovmatrix}

The clipping procedure of the negatives elements of the power spectrum $P_\nu(\vec k)$ (see section \ref{practicalimplementation}) inevitably produces a bias in the estimated covariances if their are blindly obtained using eq. \ref{covmatrixestimator}. Indeed, in doing so without caution one would conclude that the covariance matrix obtained from \texttt{COVMOS} is systematically over-estimated. This is a spurious effect due to the fact that the $k=0$ mode of the non-Gaussian density field obtained from the pipeline, related to $\left< \delta \right>$, is not rigorously null.

The reason for such a difference comes from the fact that the DC mode $P_\nu(\vec k =0)$ (when applying the pipeline eq. \ref{overalltransfo}) is likely to be negative while we impose it to be null in order to make sure that the Gaussian field has a zero mean (the whole process relies on this). This is producing an undesired non null $k=0$ mode in the power spectrum of the density field $\delta$ which introduce a correlation between the volume averaged density $\bar \delta$ with the power spectrum at all $k$ modes. Thus, the power spectrum at different wave mode being correlated with a hidden variable ($\bar \delta$), the covariance appears larger than expected. In this appendix, we describe in details this effect and propose a correction.

A Monte Carlo realisation following the presented pipeline in section \ref{pipelinecomovingspace} will be characterised by a finite total number of object $X$, which will vary from a realisation to an other. 
In principle, the variance should be given by the Poisson distribution as $\sigma_X^2 = N_p^2$, where $N_p\equiv\langle X\rangle$ is the expectation value of the number of objects in the catalogues. Thus, the expectation value of the number of objects can be obtain by multiplying the volume $L^3$ by the expected number density $\rho_0$: $N_p=\rho_0L^3$.

When computing the cross correlation $C_{XP_k} \equiv \left\langle (X-N_p)\left (P(\vec k) - \langle P(\vec k)\rangle\right ) \right\rangle$ between the number of object and the amplitude of the power spectrum at a given wave mode $k$, one can show that it is given by
\begin{equation}
    C_{XP_k} = N_p k_F^3 \bar B(0,k).
    \label{covxp}
\end{equation}
In eq. \ref{covxp}, $\bar B(k_1,k_2)$ stands for the shell average bi-spectrum, a quantity in principle expected to be null when taken at any $k_1$ or $k_2$ being $0$. This is given by the fact that the mean number density is supposed to be known, thus the mean density contrast is by definition zero and no power appears at any wave-modes $\vec k=\vec 0$.

 As anticipated, if a negative value appears at the $\vec k=\vec 0$ mode, then it is enforced to be zero. It results that the $\vec k = \vec 0$ mode of the density power spectrum is expected to be larger than $0$. The first effect of this is to provide an extra variance to the Poisson one, such that $\sigma_X^2 = N_p^2(1 + k_F^3P_\delta(\vec k=\vec 0) )$. 

The second effect is to induce a non-negligible shell average bi-spectrum $\bar B(0,k)$. Indeed, at leading order one can express $B(0,k) \simeq 2c_2c_1^2P^2(k)$ with respect to the Hermite coefficients $c_1$ and $c_2$, that are formally defined in \cite{Baratta:2019bta}. This is showing that an extra correlation is introduced between the total number of object (or the mean density) and the power spectrum. 
%
%
After having tried to predict precisely the correlation coefficient (using the Mehler expansion of the bi-spectrum) in order to remove it from the computed covariance, we found that the following empirical formula works well and allows to efficiently recover the correct the covariance 

\begin{equation}
P_\mathrm{deb}(k) = \left[\frac{N_{p}}{X}\right]^3(\hat P(k)-P_\mathrm{SN}),
\label{Pkdebiasing}
\end{equation}
where $P_\mathrm{SN} = (1/(2\pi))/\rho_0$ is the shot noise contribution to the power spectrum. 
The overall correction is just taking into account that the global effect of having more or less particles is to resize the amplitude of the power spectrum. 
At the end of the day, the shot noise corrected power spectrum that will be used for covariance matrix estimation will therefore be $P_\mathrm{deb}(k)$. 


\end{document}